\documentclass[12pt,final]{article}

\usepackage{hyperref}

\usepackage{cite}
\usepackage{amsmath}
\usepackage{amssymb}
\usepackage{amsthm}
\usepackage{mathrsfs}
\usepackage{ctable}

\usepackage{graphicx,color}

\newcommand{\CC}{\mathbb{C}}
\newcommand{\RR}{\mathbb{R}}
\newcommand{\QQ}{\mathbb{Q}}
\newcommand{\FF}{\mathbb{F}}

\newcommand{\ZZ}{\mathbb{Z}}
\newcommand{\PP}{\mathbb{P}}

\def\c{\gamma}

\def\m{\mu}

\def\p{\pi}

\def\t{\tau} 
\def\taubar{\bar \tau}
\def\tbar{\bar \tau}

\def\L{\Lambda}

\def\Tr{\tr}

\newcommand{\ex}{\operatorname{e}}

\newcommand{\tr}{\operatorname{tr}}

\newcommand{\xmod}{{\rm \;mod\;}}

\newcommand{\cP}{{\cal P}}
\newcommand{\hkt}{{\frac12{\rm K3}}}
\newcommand{\dpe}{{{\rm dP}_8}}

\newcommand{\bea}{\begin{eqnarray}} 
\newcommand{\eea}{\end{eqnarray}} 
\newcommand{\bee}{\begin{eqnarray*}} 
\newcommand{\eee}{\end{eqnarray*}} 
\newcommand{\al}{\begin{align*}} 
\newcommand{\eal}{\end{align*}} 
\newcommand{\be}{\begin{equation}} 
\newcommand{\ee}{\end{equation}} 
 
\newcommand{\bem}{\begin{pmatrix}} 
\newcommand{\eem}{\end{pmatrix}} 

\newcommand{\SL}{\operatorname{\textsl{SL}}}      

\begin{document}

\centerline{\Large Attractive Strings and Five-Branes,}
\centerline{\Large Skew-Holomorphic Jacobi Forms and Moonshine}
\bigskip
\bigskip
\centerline{Miranda C. N. Cheng$^a$, John F. R. Duncan$^b$, Sarah M. Harrison$^{c,d}$,}
\centerline{Jeffrey A. Harvey$^e$, Shamit Kachru$^{f}$, Brandon C. Rayhaun$^{f}$}
\bigskip
\bigskip
\centerline{$^a$Institute of Physics and Korteweg-de Vries Institute for Mathematics}
\centerline{University of Amsterdam, Amsterdam, the Netherlands}
\medskip
\centerline{$^b$Department of Mathematics and Computer Science}
\centerline{Emory University, Atlanta, GA 30322, USA}
\medskip
\centerline{$^c$Center for the Fundamental Laws of Nature}
\centerline{Harvard University, Cambridge, MA 02138, USA}
\medskip
\centerline{$^d$Department of Mathematics and Statistics and Department of Physics}
\centerline{McGill University, Montreal, QC, Canada}
\medskip
\centerline{$^e$Enrico Fermi Institute and Department of Physics}
\centerline{University of Chicago, Chicago IL 60637, USA}
\medskip
\centerline{$^f$Stanford Institute for Theoretical Physics}
\centerline{Stanford University, Palo Alto, CA 94305, USA}

\bigskip
\begin{abstract}

We show that certain  BPS counting functions for both fundamental strings and strings arising from fivebranes wrapping divisors in Calabi--Yau threefolds naturally give rise to skew-holomorphic Jacobi forms at rational and attractor points in the moduli space of string compactifications. For M5-branes wrapping divisors these are forms of weight negative one, and in the case of multiple M5-branes skew-holomorphic mock Jacobi forms arise. We further find that in simple examples these forms are related to skew-holomorphic (mock) Jacobi forms of weight two that play starring roles in moonshine. We discuss examples involving M5-branes on the complex projective plane, del Pezzo surfaces of degree one, and  half-K3 surfaces. For del Pezzo surfaces of degree one and certain half-K3 surfaces we find a corresponding graded (virtual) module for the degree twelve Mathieu group. This suggests a more extensive relationship between Mathieu groups and complex surfaces, and a broader role for M5-branes in the theory of Jacobi forms and moonshine.

\end{abstract}

\newpage
\tableofcontents

\section{Introduction}

Jacobi forms and mock Jacobi forms 
play important roles as counting functions governing black hole entropy in string theory. For a recent comprehensive discussion see \cite{DMZ}. They also play starring roles in studies of moonshine, as in, e.g., \cite{EOT,mum,mumcor}. 
Skew-holomorphic Jacobi forms, first introduced by Skoruppa in \cite{Skoruppa,Skoruppa_exp}, 
also play an important role in moonshine. Indeed, the weight one-half modular forms exhibiting moonshine for the Thompson group in  \cite{HR} 
can be recast as  the theta components of skew-holomorphic Jacobi forms, an observation extended in \cite{dhr} to obtain a larger family
of moonshine phenomena.  In this work we promote the idea that BPS counting functions
appearing in the theory of strings and wrapped fivebranes  at rational and attractor points provide a rich source of such objects
and suggest further new possibilities
for connections between moonshine, black holes, and BPS state counting. 

\medskip
Our first main observation is that half-BPS state counting functions for the heterotic string on $S^1$ at rational points in the Narain moduli space
lead directly to skew-holomorphic Jacobi forms. 
Our second main observation is that M5-branes wrapping divisors in Calabi--Yau threefolds, studied in e.g. \cite{MSW} as giving rise to black strings in M-theory, provide another natural source of skew-holomorphic Jacobi forms.
As discussed in \cite{Yin,Cheng, Denef, Manschot}, the modified elliptic genera counting supersymmetric states in these theories are 
non-holomorphic modular forms of a certain kind. 
We will see that at suitable moduli these functions can be specialized to
skew-holomorphic Jacobi forms.   A number of examples of such genera were
computed in a closely related setup in \cite{Klemm} (note that many of these do not satisfy the ``ampleness" assumption of \cite{MSW}).  
We will see, in several cases, 
that a 
skew-holomorphic Jacobi form or mock Jacobi form of weight $2$ which plays
a 
role in moonshine can be extracted.  We will focus on cases where either a single M5-brane is wrapped,
or two M5-branes are wrapped. Skew-holomorphic mock Jacobi forms appear in the latter case, due to the presence of bound states of single wrapped M5-branes. 

\medskip
Another important observation concerns the particular example of a single M5-brane wrapping a del Pezzo surface of degree one (i.e. $\PP^2$ blown up at eight points). As we explain in \S\ref{sec:egs:dP8}, the corresponding skew-holomorphic Jacobi form of weight $2$ admits an interpretation as a generating function for the graded dimension of  a graded virtual module for the sporadic simple group $M_{12}$.  
This suggests a non-trivial relationship between $M_{12}$ and del Pezzo surfaces, and a concrete path to begin its exploration. 
In \S\ref{sec:egs:halfK3} we give evidence that this relationship can be extended to half-K3 surfaces (i.e. blow-ups of $\PP^2$ at nine points) at certain moduli.
In addition to this, the form in which the relevant skew-holomorphic Jacobi forms are found points toward a concrete construction in terms of a vertex algebra attached to a certain indefinite lattice.

\medskip
The plan of this note is as follows. In \S\ref{sec:shjf} we give a brief review of skew-holomorphic Jacobi forms.  In \S\ref{sec:SHJFS1} we discuss $S^1$ heterotic string compactifications at rational points in Narain moduli space and highlight the connection between the BPS counting function and
skew-holomorphic Jacobi forms. In \S\ref{sec:M5EG} we review the M5-brane elliptic genus, and show that, when evaluated at a relevant attractor point in moduli space, it gives a 
skew-holomorphic 
Jacobi form of weight $-1$.  In \S\ref{sec:egs} we discuss
several examples where weight $2$ skew-holomorphic (mock) 
Jacobi forms that are implicated in 
moonshine appear. 
The discussion of $M_{12}$ and del Pezzo surfaces appears in \S\ref{sec:egs:dP8}, and this is extended to half-K3 surfaces in \S\ref{sec:egs:halfK3}. Some further details and supporting data for these relationships appears in Appendix
A.

\section{Skew-holomorphic Jacobi forms}\label{sec:shjf}

We briefly review skew-holomorphic Jacobi forms in this section, referring to \cite{DMZ} or \cite{ChengDuncan} for more details.

\medskip

In very general terms, a skew-holomorphic Jacobi form of weight $k$ and index $m$ is a function of the form
\begin{gather}\label{eqn:shjf}
	\varphi(\tau,z)=\sum_{r\xmod 2m}\overline{f_r(\tau)}\theta_{m,r}(\tau,z)
\end{gather}
where the {\em theta-coefficients} $f_r$ are the components of a holomorphic vector-valued modular form of weight $k-\frac12$.
In this work we consider $m\in \frac12\ZZ$, and use
\begin{gather}\label{eqn:thetamr}
\theta_{m,r}(\tau,z):=\sum_{\substack{\ell\in\ZZ+m\\\ell=r\xmod 2m}}\ex(m\ell)y^{\ell}q^{\frac{\ell^2}{4m}},
\end{gather}
for $r\in \ZZ+m$, where $\ex(x) := e^{2\pi i x}$ and $y:=\ex(z)$ and $q:=\ex(\tau)$. Usually it is required that $f_r(\tau)=O(1)$ as $\Im(\tau)\to \infty$, for all $r$, and the term {\em weakly skew-holomorphic} is used when this is relaxed to $f_r(\tau)=O(e^{C\Im(\tau)})$ for some $C>0$. A skew-holomorphic mock Jacobi form is a function as in (\ref{eqn:shjf}) for which the $f_r$ are mock modular forms in the usual sense (cf. e.g. \cite{DMZ}).

\medskip
In order to formulate some examples define the {\em thetanullwerte}
\begin{gather}
\begin{split}\label{eqn:thetamrj}
\theta_{m,r}^0(\tau)&:=\sum_{\substack{\ell\in\ZZ+m\\\ell=r\xmod 2m}}\ex(m\ell)q^{\frac{\ell^2}{4m}},\\ 
\theta_{m,r}^1(\tau)&:=\sum_{\substack{\ell\in\ZZ+m\\\ell=r\xmod 2m}}\ex(m\ell)\ell q^{\frac{\ell^2}{4m}}. 
\end{split}
\end{gather}
Then for $k\in \{1,2\}$ and $m\in \frac12\ZZ$ the function
\begin{gather}\label{eqn:tmj}
	t_{k,m}(\tau,z):=\sum_{r\xmod 2m}\overline{\theta^{k-1}_{m,r}(\tau)}\theta_{m,r}(\tau,z)
\end{gather}
is a skew-holomorphic Jacobi form of weight $k$ and index $m$. 
These {\em theta-type} skew-holomorphic Jacobi forms (cf. \S3.1 of \cite{ChengDuncan}) arise as shadows in umbral moonshine. For example, if 
\begin{gather}
	H^{(2)}(\tau)=-2q^{-\frac18}+90q^{\frac78}+462q^{\frac{15}8}+1540q^{\frac{23}8}+\dots
\end{gather} 
is the McKay--Thompson series attached to the identity element of $M_{24}$ by Mathieu moonshine \cite{EOT} then $\phi^{(2)}(\tau,z):=H^{(2)}(\tau)(\theta_{2,-1}(\tau,z)-\theta_{2,1}(\tau,z))$ is a (weakly holomorphic) mock Jacobi form of weight $1$ and index $2$, and its shadow is proportional to $t_{2,2}(\tau,z)$.

\medskip
The half-integral index theta series (\ref{eqn:thetamr}), (\ref{eqn:thetamrj}) include some familiar examples, which will play a role in \S\ref{sec:egs}. For instance, for $m=\frac12$ we have
\begin{gather}
\begin{split}\label{eqn:thetahalfhalf}
\theta_{\frac12,\frac12}(\tau,z)&=i\sum_{n\in \ZZ}y^{n+\frac12}q^{\frac12(n+\frac12)^2}\\
&=iy^{\frac12}q^{\frac18}\prod_{n>0}(1-y^{-1}q^n)(1-yq^n)(1-q^n).
\end{split}
\end{gather}
So $\theta_{\frac12,\frac12}^0$ vanishes identically, but $\theta_{\frac12,\frac12}^1=i\eta^3$, where $\eta$ denotes the Dedekind eta function, $\eta(\tau):=q^{\frac1{24}}\prod_{n>0}(1-q^n)$. For $m=\frac32$ we have $\theta_{\frac32,\frac32}^0=0$ and 
\begin{gather}\label{eqn:thetathreehalvespmhalf}
	\theta_{\frac32,\pm \frac12}^0(\tau)
	=\mp i\sum_{n\in \ZZ}(-1)^nq^{\frac1{6}(3n\pm \frac12)^2}
	=\mp i \eta(\tau).
\end{gather}
Also note the identity $t_{2,\frac12}(\tau,z)=\frac12t_{2,2}(\tau,\frac12z)$, which hints at an index $m=\frac12$ formulation of Mathieu moonshine. A broader context for this is given in \cite{ChengDuncanN2}.

\medskip
From a number theoretic point of view skew-holomorphic Jacobi forms 
play a complementary role to holomorphic Jacobi forms in a particular formulation of the Shimura correspondence, developed by Skoruppa and Zagier \cite{SkoruppaZagier,Skoruppa_exp,Skoruppa_Hee}. 
Consequently there are Waldspurger-type results relating Fourier coefficients of holomorphic and (non theta-type) skew-holomorphic Jacobi forms of weight at least 2 to special values of $L$-functions of cuspidal modular forms with level (cf. \cite{Skoruppa_Hee}). This mechanism plays an important role in the arithmetic geometry of elliptic curves according to the celebrated Birch--Swinnerton-Dyer conjecture. Applications to moonshine have appeared, for instance, in \cite{ChengDuncan} and \cite{DMO}.

\medskip
Our focus in \S\ref{sec:egs} will be on examples of M5-brane configurations that produce (weakly) skew-holomorphic (mock) Jacobi forms of weight $2$.

\section{Rational heterotic string compactifications}\label{sec:SHJFS1}

In this section we analyze examples of $S^1$ compactifications of the heterotic string at points in the Narain moduli space that 
correspond to rational conformal field theories. By definition these are points at which there is an extended chiral algebra with the CFT containing
a finite number of irreducible representations of the chiral algebra.  The partition function thus decomposes into a finite sum of the form
\begin{align}
Z(q) = \sum_{j,\bar{j}} N_{j\bar{j}}\chi_j(q)\bar{\chi}_{\bar{j}}(\bar{q})
\end{align}
where the $N_{j\bar{j}}$ are non-negative integers and the $\chi_j$ ($\bar{\chi}_{\bar{j}}$) furnish holomorphic (anti-holomorphic) irreducible characters of the extended chiral algebra which is larger than the Virasoro algebra. Of course, the $\chi_j$ and $\bar{\chi}_{\bar{j}}$ are in general reducible with respect to the Virasoro algebra and decompose into a possibly infinite sum of its irreducible characters. We show that the half-BPS state counting functions which arise can be written in terms of skew-holomorphic Jacobi forms. See \cite{moorearith} for a general discussion of the relationship between rational CFT and attractor points
in the moduli space of string compactifications.

\subsection{The rational Gaussian model}\label{sec:rationalGaussian}

The $c=1$ Gaussian model,  corresponding to string compactification on a $S^1$ of radius $R$, is defined by an embedding
of the unique unimodular even lattice of signature $(1,1)$ into $\RR^{1,1}$. We denote the embedded lattice by $\Gamma^{1,1}$ and write lattice vectors and their standard projections as $p=(p_L,p_R)$. More generally, for $r= s \xmod 8$, we will use $\Gamma^{r,s}$ to denote an embedding of the unique unimodular even lattice of signature $(r,s)$ into $\RR^{r,s}$. Using conventions
in which the inverse string tension is $\alpha'=2$ we have
\bea
p_L &= \frac{n}{R} + \frac{w R}{2} \\
p_R &= \frac{n}{R} - \frac{w R}{2}
\eea
with $n,w \in \ZZ$. 
The moduli space of the $c=1$ Gaussian model is 
\be
\ZZ_2 \backslash O(1,1;\RR)/O(1) \times O(1) \simeq \ZZ_2 \backslash \RR_+
\ee
where the $\ZZ_2$ acts as T-duality, $R \mapsto \frac{2}{R}$. Thus the moduli space is the half line $[\sqrt{2}, \infty )$ parametrized by $R$.

\medskip
The model contains holomorphic  and anti-holomorphic $U(1)$ currents $J$, $\bar J$  with eigenvalues proportional to $p_L,p_R$.
Introducing  chemical potentials $\zeta=(\zeta_L, \zeta_R)$ to keep track of these
$U(1)$ charges leads to
the partition function
\be
Z(\tau,\zeta) = 
\Theta_{1,1}(R;\tau,\zeta)
|\eta(\tau)|^{-2}
\ee
where
\be
\Theta_{1,1}(R;\tau,\zeta) := \sum_{p \in \Gamma^{1,1}} q^{\frac12p_L^2}  {\bar q}^{\frac12p_R^2} e^{2 \pi i \zeta \cdot p} \, .
\ee

\medskip
Let $\Gamma_R := \{(0,p_R)\in \Gamma^{1,1}\}$ be the lattice of right-moving momenta. We now consider rational points in the moduli space where $R^2 \in \QQ$, and say that
$\Gamma_R$ is generated by $p_0$.  
In order to facilitate the comparison to skew-holomorphic Jacobi forms using the conventions of the previous section we specialize to the case
$\zeta(z) = \bar{z} p_0$ (this corresponds to choosing the normalization of $\bar J$ such that the associated charge has integer eigenvalues).  We
will show that the Siegel--Narain theta function $\Theta_{1,1}$ is the complex conjugate of a weight one skew-holomorphic Jacobi form of theta-type at
such rational points.

\medskip
Consider first the self-dual point $R=\sqrt{2}$. We then have
\be
\Theta_{1,1}(\sqrt{2};\tau,\zeta(z))= \sum_{n,w \in \ZZ} q^{\frac{(n+w)^2}4}{\bar q}^{\frac{(n-w)^2}4}  {\bar y}^{n-w} 
\ee
with $\bar y=\ex(- \bar z)$. 
Breaking the sum into terms with $n+w$ even and $n+w$ odd gives
\be \label{thoneone}
\Theta_{1,1}(\sqrt{2};\tau,\zeta(z)) = \sum_{r \xmod 2} \overline{ \theta_{1,r}(\tau,z)} \theta^0_{1,r}(\tau) = \overline{ t_{1,1}(\tau, z)} \, ,
\ee
which is of the claimed form.  

\medskip
It is not difficult to generalize this to general rational $\frac{R^2}{2}$, a problem which appears as Exercise 10.21 in \cite{yellowbook}. We write $R^2= 2 \frac{\kappa'}{\kappa}$ with $\kappa',\kappa$ coprime integers. We
then have
\begin{gather}
\begin{split}
\Theta_{1,1}\left(\sqrt{2\tfrac{\kappa'}\kappa};\tau, \zeta(z)\right) & = \sum_{(p_L,p_R) \in \Gamma^{1,1}} q^{\frac12p_L^2}\bar{q}^{\frac12p_R^2}{\bar y}^{p_R\sqrt{2\frac{\kappa'}\kappa}} \\
&= \sum_{n,w}q^{\frac{(n\kappa+w\kappa')^2}{4\kappa\kappa'}}\bar{q}^{\frac{(n\kappa-w\kappa')^2}{4\kappa\kappa'}} {\bar y}^{n\kappa-w\kappa'}.
\end{split}
\end{gather}
Now define $r_0,s_0$ to be integers for which $\kappa r_0-\kappa's_0 = 1$, which is always possible since $\kappa,\kappa'$ are coprime. Define $\omega_0$
and $r$ to be the values of  $\kappa r_0+\kappa's_0$ and $n\kappa+w\kappa' $ modulo $2\kappa\kappa'$ respectively.  Then a short computation shows that
$n\kappa-w \kappa'= \omega_0 r ~{\rm mod}~2\kappa\kappa'$ which allows us to write
\begin{align}
\Theta_{1,1}\left(\sqrt{2\tfrac{\kappa'}\kappa};\tau,\zeta(z)\right) &= \sum_{r\xmod 2\kappa\kappa'}\sum_{n= \omega_0 r\xmod 2\kappa\kappa'}\bar{q}^{\frac{n^2}{4\kappa\kappa'}} {\bar y}^n\sum_{m= r\xmod 2\kappa\kappa'}q^{\frac{m^2}{4\kappa\kappa'}} \\
&= \sum_{r\xmod 2\kappa\kappa'}\overline{\theta_{\kappa\kappa',\omega_0 r}(\tau,z)} \theta^0_{\kappa\kappa',r}(\tau)\nonumber
\end{align}
This is almost of the desired form except for the factor of $\omega_0$. This factor can be understood in terms of an automorphism of the fusion rule algebra as discussed in \cite{moorearith, MooreSeiberg} and in the mathematical literature is related to well-known
objects, namely the Eichler--Zagier operators which played a prominent role in \cite{mum}.

\medskip
To see this, we can perform a trivial rewriting of the previous equation,
\begin{align} \label{skewez}
\Theta_{1,1}\left(\sqrt{2\tfrac{\kappa'}{\kappa}};\tau,\zeta(z)\right) &= \sum_{s,r\xmod 2\kappa\kappa'}\overline{\theta_{\kappa\kappa',s}(\tau, z)}\delta_{s,\omega_0r}\theta^0_{\kappa\kappa',r}(\tau) \, .
\end{align}
The matrix with matrix elements $\delta_{s,\omega_0 r}$ is an Eichler--Zagier matrix,
\begin{align} 
\Omega_{\kappa\kappa'}(\kappa)_{sr} = \delta_{s,\omega_0 r} \, ,
\end{align}
see \cite{mum} for conventions. Recall that $\Omega_m(n)_{sr} = 1$ if $s+r = 0\xmod 2n$ and $s-r= 0 \xmod \frac{2m}{n}$, and 0 otherwise. An easy calculation shows that the two conditions required for a matrix element of $\Omega_{\kappa\kappa'}(\kappa)$ to be nonzero are equivalent to $s= \omega_0 r\xmod 2\kappa\kappa'$:
\begin{align}
 (s-\omega_0r)\xmod 2\kappa &= s-(\kappa r_0+\kappa's_0)r \xmod 2\kappa \\
&= s + (\kappa r_0-\kappa's_0)r\xmod 2\kappa \nonumber \\
&= (s +r) \xmod 2\kappa \nonumber \\
(s-\omega_0r) \xmod  2\kappa' & = s-(\kappa r_0+\kappa's_0)r \xmod 2\kappa' \\
&= s-(\kappa r_0-\kappa's_0)r \xmod 2\kappa' \nonumber \\
&= (s-r)\xmod 2\kappa' \nonumber 
\end{align}
from which it easily follows that $\Omega_{\kappa\kappa'}(\kappa)_{sr}=\delta_{s,\omega_0r}$ and thus that (\ref{skewez}) is the complex conjugate of a skew-holomorphic
Jacobi form:
\begin{align}
\Theta_{1,1}\left(\sqrt{2\tfrac{\kappa'}{\kappa}};\tau,\zeta(z)\right) &= \overline{\theta_{\kappa\kappa'}(\tau,z)}\cdot \Omega_{\kappa\kappa'}(\kappa)\cdot \theta^0_{\kappa\kappa'}(\tau) \,
\end{align}
where we have suppressed the vector indices in the above equation.

\subsection{Heterotic strings with Wilson lines}

We now explain the relevance of this computation to BPS state counting for heterotic strings on $S^1$. In this case the Narain
moduli space has dimension $17$, corresponding to the radius of the $S^1$ and a choice of Wilson lines in the Cartan subalgebra
of $E_8 \times E_8$ or $\mathrm{Spin}(32)/\ZZ_2$. Half-BPS states correspond to right-moving ground states with arbitrary left-moving
excitations \cite{dh} and have squared mass proportional to $p_R^2$. The generating function for these BPS states, summed over
all $p_R^2$ and weighted by a chemical potential for $p_R$ is given by\footnote{For the purpose of comparing to black hole microstate counts, we comment that the partition function defined here has the same leading asymptotic behavior as the familiar $1/\eta^{24}(\tau)$, receiving only subleading corrections from the theta function. Similar comments apply to the rest of the counting functions considered in this paper.}
\be
Z_{\mathrm{BPS}}(\tau,\zeta) =  \Theta_{17,1}(\tau,\zeta){\eta^{-24}(\tau)}
\ee
where now
\be
\Theta_{17,1}(\tau,\zeta) := \sum_{p \in \Gamma^{17,1}} q^{\frac12p_L^2}\bar q^{\frac12p_R^2} e^{2\pi i \zeta\cdot p}  \, .
\ee
We expect that $Z_{\mathrm{BPS}}$ can be written in terms of skew-holomorphic Jacobi forms at rational points in the Narain
moduli space 
\be \label{narain}
{\cal N}_{17,1}:=O(17,1,\ZZ) \backslash O(17,1,\RR)/O(17) \times O(1).
\ee
We will show this explicitly for two examples below, and defer comments about the general case to \S\ref{sec:rationaltoroidal}.

\medskip
The first example involves considering points in the moduli space (\ref{narain}) where the Wilson lines are turned off. At these points, the embedded lattice $\Gamma^{17,1}$ respects the standard splitting $\RR^{17,1} = \RR^{16}\oplus \RR^{1,1}$ in the sense that $L := \Gamma^{17,1}\cap \RR^{16}$ is a positive-definite even unimodular lattice with rank 16 and $\Gamma^{17,1}\cap \RR^{1,1}$ is unimodular and even with signature (1,1). If we further specialize to points in the moduli space where the $\Gamma^{1,1}$ corresponds to a rational CFT of radius $R = \sqrt{2\tfrac{\kappa'}{\kappa}}$ we then find
\begin{gather}
\begin{split}
Z_{\mathrm{BPS}}(\tau,\zeta(z)) &= {\Theta_{17,1}(\tau, \zeta(z))}{\eta^{-24}(\tau)} \\
&= \sum_{(p_L,p_R) \in \Gamma^{17,1}}q^{\frac12p_L^2}\bar{q}^{\frac12p_R^2} {\bar y}^{p_R\sqrt{2\kappa'\kappa}} \\
&= \overline{\theta_{\kappa\kappa'}(\tau,z)} \cdot\Omega_{\kappa\kappa'}(\kappa) \cdot\theta^0_{\kappa\kappa'}(\tau) {\Theta_L(\tau)}{\eta^{-24}(\tau)}
\end{split}
\end{gather}
where $\Theta_L$ is the theta-function attached to the lattice $L$. There are only two even unimodular lattices of rank $16$; namely $E_8\oplus E_8$ and $D_{16}^+$. In both cases $\Theta_L$ is the unique weight $8$ modular form that satisfies $\Theta_L(\tau) = 1+O(q)$ as $\Im(\tau) \to \infty$. The partition function
$Z_{\mathrm{BPS}}(\tau,\zeta(z))$ is the complex conjugate of a weakly skew-holomorphic Jacobi form of weight $-3$.

\medskip
We can obtain a more subtle rational point by utilizing a construction due to Nikulin \cite{nikulin}. Let $\Lambda_L$ be an even, rank $7$, positive-definite lattice
which is primitively embedded into an even, unimodular, rank 24 lattice $N$, thus $N$ is the Leech lattice or one of the $23$ Niemeier lattices.
Let $\Lambda_R$ be a negative-definite lattice bijectively isometric to $\Lambda_L$ (up to an overall minus sign in the quadratic form) and primitively embedded as a sublattice of the (negative-definite) $E_8$ root lattice. 
Define $\Gamma_L:=\Lambda_L^\perp\cap N$ and $\Gamma_R := \Lambda_R^\perp\cap (-E_8)$ to be the orthogonal complements of $\Lambda_L$ and $\Lambda_R$ respectively. 
Then the lattice
\be
\Gamma := \bigoplus_i\left( (\Gamma_L,0)\oplus (0, \Gamma_R) + (g_L^{(i)}, g_R^{(i)}) \right)
\ee
is an even, unimodular lattice of signature $(17,1)$. In the above $g_{L}^{(i)}$ and $g_R^{(i)}$ are glue vectors which run over the non-trivial elements
of the discriminant group of $\Gamma_L, \Gamma_R$, such that the map $g_R^{(i)}\to g_L^{(i)}$ is an isometry. 

\medskip
Our second example will use this construction for the choice $\Lambda_L = E_7$ which is embedded in the Niemeier lattice with root system $A_{17}E_7$, and  $\Lambda_R = (-E_7)$  which is primitively embedded into $(-E_8)$.\footnote{For details on the construction of such lattices, see e.g. \cite{CS,Ebeling}.}
Our conventions for the $A_{17}E_7$ root lattice are as follows.  Take $e_1, e_2, \dots, e_{26}$ to be an orthonormal basis for $\RR^{26}=\RR^8 \oplus \RR^{18}$ and take the $E_7$ root lattice to be embedded in the first $\RR^8$
with simple roots $r_i:=e_{i+2}-e_{i+1}$ for $i=1,2, \dots, 6$, and $r_7 := \frac12(e_1+e_2+e_3+e_4-e_5-e_6-e_7-e_8)$. For the $A_{17}$ root
system we embed in the $\RR^{18}$ factor  and take the simple roots to be $r_i := e_i-e_{i+1}$ for $i=9, 10, \dots, 26$.
Recall the  construction of  the Niemeier lattice $N$ corresponding to the $A_{17}E_7$ root system. Letting $L^\star$ denote the dual of a lattice $L$, we have embeddings 
\begin{align}
A_{17}E_7 \subset N \subset (A_{17}E_7)^\star
\end{align}
which implies that $N/A_{17}E_7$ is a subgroup of $(A_{17}E_7)^\star/A_{17}E_7$. Moreover, since $N$ is an even lattice,  it is an isotropic subgroup, meaning that the quadratic form of the discriminant group restricted to $N/A_{17}E_7$ vanishes.  Now, the discriminant group of $A_{17}E_7$ is $\ZZ_{2}\times \ZZ_{18}$ from  the $E_7$ and $A_{17}$ factors respectively. 

\medskip
The $\ZZ_2$ component of the discriminant group is generated by 
\begin{align}
v:=\frac{1}{4}(3 e_1-e_2-e_3-e_4-e_5-e_6+3 e_7-e_8),
\end{align}
while the $\ZZ_{18}$ is generated by 
\begin{align}
w:= \frac{1}{18}\left(17 e_9 - \sum_{i=10}^{26} e_i\right) .
\end{align}
One can check that the quadratic form on the discriminant group vanishes on the isotropic subgroup 
$\langle v+3w\rangle \simeq \ZZ_6$.
The Niemeier lattice is obtained as 
\begin{align}
N = \bigcup_{n=0,\dots,5} (A_{17}E_7 + n(v+3w))
\end{align}
and the orthogonal complement of the $E_7$ root lattice in $N$ is easily seen to be
\begin{align}
\Gamma_L = A_{17} \cup ( A_{17}+6w )\cup ( A_{17}+12w ).
\end{align}
One can convince oneself that the discriminant group $\Gamma_L^\star/\Gamma_L = \langle 3w\rangle\simeq \ZZ_2 $. 

\medskip
The even unimodular lattice $\Gamma$ that we obtain in this way satisfies 
\begin{align}
\Theta_{17,1}(\tau,\zeta(z)) = \Theta_{\Gamma_L}(\tau)\overline{\theta_{1,0}(\tau,z)} + \Theta_{\Gamma_L + 3w}(\tau)\overline{\theta_{1,1}(\tau,z)}.
\end{align}
and an explicit computation of the theta coefficients yields
\begin{align}
\Theta_{\Gamma_L}(\tau) &= 1 + 306q + 55488q^2 + 1161984q^3 + 10054242q^4 \\
& \hspace{.2in} + 53585088q^5 + 210351744q^6 + 668519424q^7+ \cdots ,\nonumber \\
\Theta_{\Gamma_L+3w}(\tau) &= 1632q^{5/4} + 134912q^{9/4} + 2110176q^{13/4} \\
& \hspace{.2in} + 15898368q^{17/4} + 76968384q^{21/4} + 286866432q^{25/4} + \cdots.  \nonumber
\end{align}

\medskip
We can identify the above theta coefficients further using results in \cite{Skoruppa}. Skoruppa classifies the weight $k$ index 1 skew-holomorphic Jacobi forms:
\begin{align}
J^{\mathrm{sk}}_{k,1} = M_{k-1}(\mathrm{SL}_2(\ZZ))\cdot t_{1,1}(\tau,z) \oplus M_{k-3}(\mathrm{SL}_2(\ZZ))\cdot U(\tau,z).
\end{align}
Here, $M_k(\mathrm{SL}_2(\ZZ))$ is the space of weight $k$ holomorphic modular forms for $\mathrm{SL}_2(\ZZ)$, and 
\begin{align}
U(\tau,z) &:= \frac{12}{\pi i}\frac{\partial}{\partial\bar\tau} t_{1,1}(\tau,z) + \overline{E_2(\tau)}t_{1,1}(\tau,z), \\
E_2(\tau) &:= 1 - 24\sum_{\ell\geq 1}\Big(\sum_{d\mid \ell} d\Big) q^\ell.
\end{align}

\medskip
Letting $E_k(\tau) = 1 + O(q)$ be the Eisenstein series of weight $k$, the weight nine skew-holomorphic forms (and in particular the function we found above) should be of the form 
\begin{align}
a \overline{E_4(\tau)^2}t_{1,1}(\tau,z) + b \overline{E_6(\tau)}U(\tau,z).
\end{align}
One can verify that the theta function we computed earlier corresponds to the choice $a=\frac56$ and $b=\frac16$,
\begin{align}
\Theta_{17,1}(\tau,\zeta(z)) &= \frac{1}{6}\left(5\overline{E_4(\tau)^2}t_{1,1}(\tau,z) + \overline{E_6(\tau)}U(\tau,z)\right).
\end{align}
We are then left with the BPS counting function
\be
Z_{\mathrm{BPS}}(\tau,\zeta(z))= \frac{1}{6}\left(5 E_4(\tau)^2 \overline{t_{1,1}(\tau,z)} + E_6(\tau)\overline{U(\tau,z)}\right)\eta^{-24}(\tau).
\ee

\medskip
It should not be hard to generalize this analysis to other rational points in the moduli space (\ref{narain}) at which the BPS state
counting function can be expressed in terms of skew-holomorphic Jacobi forms. 

\subsection{Rational toroidal compactifications}\label{sec:rationaltoroidal}

As a technical aside, we would like to briefly sketch the general construction which underlies the examples of the previous sections. Quite generally, toroidal string compactifications correspond to Narain lattices $\Gamma$ of signature $(d+8s,d)$. The points in the moduli space of such lattices where the associated CFT becomes rational are specified by triples $(\Gamma_L,\Gamma_R,\phi)$, where we demand that $\phi:\Gamma_R^\star/\Gamma_R\to\Gamma_L^\star/\Gamma_L$ be an isometric bijection of the discriminant groups. The discriminant groups $\Gamma_{L}^\star/\Gamma_{L}$ and $\Gamma_R^\star/\Gamma_R$ inherit their norms from the norms on $\Gamma_{L}^\star$ and $\Gamma_R^\star$ reduced modulo 2. Using $\phi$ to obtain so-called {glue vectors} $(\phi(\lambda),\lambda)$, we may construct the full, rational, unimodular lattice from this data as 
\begin{align}
\Gamma:= \bigcup_{\lambda \in \Gamma_R^\star/\Gamma_R} \Big(\Gamma_L\oplus \Gamma_R + (\phi(\lambda),\lambda)\Big).
\end{align}
It easily follows that the Siegel--Narain theta function admits the decomposition 
\begin{align}
\Theta_\Gamma(\tau)  = \sum_{\lambda \in \Gamma_R^\star/\Gamma_R} \Theta_{\Gamma_L+\phi(\lambda)}(\tau)\overline{\Theta_{\Gamma_R+\lambda}(\tau)}
\end{align}
where we have defined 
\begin{align}
\Theta_{L+\lambda}(\tau) := \sum_{\gamma \in L + \lambda}q^{\frac{\gamma^2}2}
\end{align}
for an arbitrary positive-definite, even lattice $L$. In this construction, $\Gamma_L := \{(p_L,0)\in \Gamma\}$ is the lattice of purely left-moving momenta, and similarly for $\Gamma_R$. See \S 10.2 of \cite{moorearith} for a more detailed discussion. 

\medskip
In the previous sections, we specialized to $d=1$ and exploited the fact that the right-moving momentum lattice must be of the form $\Gamma_R \simeq \sqrt{2m}\ZZ$ with associated theta-function 
\begin{align}
\Theta_{\sqrt{2m}\ZZ+r}(\tau) = \theta^0_{m,r}(\tau)
\end{align}
for $r \text{ in }\Gamma_R^\star/\Gamma_R \simeq \ZZ_{2m}$. Indeed, upon flavoring by an additional $\bar{J}$ quantum number, we find that the points in the moduli space with $\Gamma_R \simeq \sqrt{2m}\ZZ$ recovered (complex conjugates of) index $m$ skew-holomorphic Jacobi forms. 

\medskip
In this language, the $c=1$ Gaussian model with radius $R = \sqrt{2\tfrac{\kappa'}{\kappa}}$ corresponds to the triple $(\sqrt{2\kappa'\kappa}\ZZ,\sqrt{2\kappa'\kappa}\ZZ,r\to\omega_0 r)$, with the gluing of left and right-moving momentum lattices specified by the isometry ``multiplication by $\omega_0$.'' The different choices of isometries give rise to different Eichler--Zagier matrices $\Omega_{\kappa\kappa'}(\kappa)$ which commute with the action of the modular group on the thetanullwerte. Similar comments should apply to the problem of classifying the rational points in the moduli space of the heterotic string with Wilson lines, as well as the skew-holomorphic Jacobi forms which arise. 

\medskip
We now turn to a richer source of strings---those arising from wrapped M5-branes---and show that their associated elliptic genera
can also be expressed in terms of skew-holomorphic Jacobi forms. 

\section{The M5-brane elliptic genus}\label{sec:M5EG}

Here we review basic facts about the worldsheet theory on a wrapped M5-brane.

\subsection{Multiplets}\label{sec:M5EG:mult}

The M5-brane wrapping a divisor in a Calabi--Yau threefold gives rise, at low-energies, to an effective string, sometimes called an
``MSW string,"
with (0,4) worldsheet supersymmetry.  This theory was studied in detail from various viewpoints in, e.g.,
\cite{MSW,Yin,Cheng,Moore}.
 Suppose the M5-brane is wrapping a divisor ${\cal P}$ in a Calabi--Yau
threefold $X$.  Then the low-energy theory on the effective string (arising from dimensional reduction of the M5-brane
worldvolume fields) is as follows.

\medskip
Consider the inclusion map 
$$ i:{\cal P} \to X~.$$
This naturally gives rise to a pullback map $i^*:H^2(X,\ZZ)\to H^2({\cal P},\ZZ)$. 
 We define $\Lambda$ to be $i^*(H^2(X,{\mathbb Z}))$ equipped with the bilinear form given by 
 $(A|B):=-\int_{{\cal P}} A\wedge B$. 
 The pullback two-forms
$i^* \alpha \in H^2({\cal P},{\mathbb Z})$ can be associated with chiral worldvolume fields in 
the $(0,4)$ worldsheet $\sigma$-model as follows.

\medskip
\noindent$\bullet$
Self-dual two-forms on ${\cal P}$ that extend non-trivially to $X$ give rise to left-moving scalars on the
worldsheet.   

\medskip
\noindent
$\bullet$ Anti self-dual two-forms on ${\cal P}$ that extend non-trivially to $X$ give rise to right-moving scalars
on the worldsheet. 

\medskip
In fact, for a Calabi--Yau threefold $X$, the K\"ahler form is the only two-form that pulls back to an anti self-dual form on ${\cal P}$. As a result there are $b^{2}(X) - 1$ left-moving scalars and 1 right-moving scalar coming
from these sources.

\medskip
It is important to remember that 
the worldsheet fields include universal worldsheet multiplets arising from (super) Goldstone modes.  This gives three additional non-chiral scalars that can translate the effective string.  The total of four right-moving bosons (including the one arising from the pullback of the K\"ahler form) have four Fermi superpartners arising from the (0,4) supersymmetry.  Zero modes of these fermions lead to a modification of the definition of the M5-brane
ellliptic genus relative to the conventional elliptic genus (see (\ref{part})) as the conventional quantity would vanish in this circumstance.  

\medskip
In a model-dependent way, there are also additional fields present in the generic wrapped M5-brane theory.  These parametrize the moduli space of motions
of the wrapped divisor in the Calabi--Yau space $X$.  
Although our subsequent discussion will be independent of these fields it should be mentioned that in the limit of large central charge, where the effective string can sometimes be related to a weakly curved black string, they constitute the most numerous degrees of freedom.

\subsection{The index}\label{sec:M5EG:index}

For a fixed M5 theory the worldsheet elliptic genus can be defined as follows.
First define
\be \label{part}
Z'(\t,\zeta):= \Tr_R\left( F^2 (-1)^F e^{\pi ip\cdot Q}
e^{2\pi i \tau \left( L_0-\frac{c_L}{24}\right)}e^{-2\pi i \taubar
\left( \bar{L}_0-\frac{c_R}{24}\right)} e^{2\pi i \zeta\cdot Q} \right). 
\;
\ee
Here the $\zeta^a$ are chemical potentials and the $Q_a$ are charges under the $b_2(X)$ abelian currents
associated with the chiral bosons; i.e. $(b_2 - 1)$ left-moving currents, and a single right-moving current. 
The $p^a$ parametrize the (discrete) choice of divisor in $H_4(X,{\mathbb Z})$ that the M5-brane wraps.
The fermion number is defined in the usual way as twice the charge of the $U(1)$ generator in the $SU(2)_R$
R-symmetry which exists in the ${\cal N}=4$ superconformal algebra.  The extra factor of $F^2$ as compared
to the conventional elliptic genus is present in order to absorb the fermion zero modes mentioned above.

\medskip
This quantity isn't quite the one we want to work with, as it includes information about the momenta in the ${\mathbb R}^3$ transverse to the effective string in the non-compact directions of space.
Instead, the {\em generalized elliptic genus} $Z(\tau,\zeta)$ is defined by requiring that
\be \label{mod5}
Z'(\t,\zeta) = Z(\t,\zeta)\,\int d^3 \vec{\p} \,(e^{2\pi i \tau} e^{-2\pi i \bar{\tau}} )^{\frac{1}{2} \vec{\p}^2} =  Z(\t,\zeta) \,(2 \,{\rm Im} \tau)^{\frac32} .
\ee

\medskip
It is easy to see \cite{Yin,Cheng,Denef} that $Z$ has weight $\left(-\frac{3}{2},\frac{1}{2}\right)$, in the sense that
\begin{gather}
Z\left(\frac{a\tau+b}{c\tau+d},\frac{\zeta}{c\tau + d}\right)\frac{(c\tau + d)^2}{|c\tau+d|}\ex\left(m\frac{cz^2}{c\bar\tau+d}\right) 
=  \chi\left(\begin{smallmatrix}a&b\\c&d\end{smallmatrix}\right)Z(\tau,\zeta)
\end{gather}
for some $m$, and some multiplier $\chi:\SL_2(\ZZ)\to \CC$, when $\left(\begin{smallmatrix}a&b\\c&d\end{smallmatrix}\right)\in\SL_2(\ZZ)$. 
This is what we would expect from (the complex conjugate of) a skew-holomorphic Jacobi form of weight $-1$ (with a multiplier system).
The shift in charges under a large gauge transformation states that 
the generalized index admits a 
decomposition
\begin{align}\label{decomposetheta}
Z(\t,\zeta)&= \sum_{\m \in \L^*/\L} \,\Theta_\m(\t,\zeta) \,h_\m(\t)
\end{align}
into Siegel--Narain theta functions
\begin{align}\label{eqn:SieNartheta}
\Theta_\m(\t,\zeta)&:= \sum_{Q\in \m+\L+\frac{p}{2}} \, 
{\ex} \left(\tfrac{\t}{2}(Q_+|Q_+)+\tfrac{\tbar}{2}(Q_-|Q_-)+(Q|\zeta+\tfrac{p}{2})\right),
\end{align}
where the splitting $Q=Q_++Q_-$ depends on the Grassmannian $\frac{O(b_2-1,1)}{O(b_2-1)\times O(1)}$.
In the
sequel we will always set the chemical potentials conjugate to the left-moving currents to zero, and only keep
track of the right-moving chemical potential.

 \medskip
 Next we will show that, at certain points in the moduli space, the generalized elliptic genus of an MSW string is naturally a skew-holomorphic (mock) Jacobi form.
 First, recall that in the presence of an MSW string the Calabi--Yau moduli which are vector multiplets in the low-energy supergravity undergo an ``attractor flow."
 That is, they flow to certain specific values at the horizon of the related black string, independent of their values at infinity in ${\mathbb R}^5$.   
 This `attractor mechanism' gives a natural preferred choice of moduli.
 In M-theory
 on $X$, the vector multiplet moduli are
 the K\"ahler moduli of $X$ (excepting the overall volume, which transforms in a hypermultiplet).
At the attractor point in moduli space, the K\"ahler form $J$ on $X$ satisfies $J\sim p$.  
As a result one can find the right-moving chiral $U(1)$ current and its associated charge to be
\be
\label{hugepig}
Q_- = \frac{p\cdot Q}{p^2} p ~.
\ee

\medskip
As we already know that $Z(\t,\zeta)$ transforms as a weight $(-\frac32,\frac12)$ modular form, what remains to check is that $\overline{Z(\t,\zeta)}$, for a specific choice of $\zeta=\zeta(z)$, satisfies the elliptic transformation
\begin{gather}
\overline{Z(\tau,\zeta(z+\lambda\tau+\mu))}\ex\left(m(\lambda^2\tau+2\lambda z + \lambda + \mu) \right)=\overline{Z(\tau,\zeta(z))}
\end{gather}
for $\lambda,\mu\in\ZZ$. This will imply that $\overline{Z}$ admits a decomposition as in (\ref{eqn:shjf}).

\medskip
Let $\zeta = \bar z p$.  At the attractor moduli the Siegel--Narain theta function $\Theta_\m(\t,\bar\t,\zeta)$    becomes 
\be
\tilde \theta_\m(\t,\bar z): = \sum_{Q\in \m+\L+\frac{p}{2}} \, (-1)^{p\cdot Q}
q^{\frac{1}{2}Q_+^2} \bar q^{\frac{1}{2} \frac{(p\cdot Q)^2}{p^2}} 
e^{2\pi i  \bar z p\cdot Q}
\ee
where $q=\ex(\tau)$ and $\bar{q}=\ex(-\bar\tau)$, and we used (\ref{hugepig}) in writing the power of $\bar q$.
We can show that 
\be
\tilde \theta_\m(\t,\bar z+n\bar\tau + m) =(-1)^{p^2 (m+n)}\ex(\tfrac12p^2 n^2) \tilde \theta_\m(\t,\bar z) 
\ee
by a shift $Q\mapsto Q+ p n$ in the sum. This verifies that, at the attractor point in moduli space, $\overline{Z(\t,p \bar z)}$ is a skew-holomorphic Jacobi form of index $\frac12{p^2}$ with elliptic variable $z$.  

\medskip
An interesting question for future work would be to determine if there are other (non-attractor) moduli where the M5-brane elliptic genus reduces to a skew-holomorphic Jacobi form.

\section{Examples}\label{sec:egs}

We now discuss several examples of M5-brane elliptic genera computed in \cite{Klemm}.  In each case we 
find a natural relation to a weakly skew-holomorphic 
Jacobi form of weight $2$ that plays a role in a moonshine.

\subsection{The projective plane}\label{sec:egs:P2}

The 
elliptic genus for one M5-brane wrapping $\PP^2$ can be written as 
\begin{gather}
	Z^{(1)}_{\PP^2}(\tau,z) = {(-i)\theta_{\frac12,\frac12}(-\bar{\tau},-{z})}{\eta^{-3}(\tau)}
\end{gather}
(cf. (\ref{eqn:thetahalfhalf})) thanks to work of G\"ottsche \cite{Gottsche}.
In comparison with \S\ref{sec:M5EG} we have kept only the chemical potential for the right-moving $U(1)$ charge, which we henceforth denote by $z$.

\medskip
So the function $\overline{Z^{(1)}_{\PP^2}(\tau,z)}$ is a skew-holomorphic Jacobi form of weight $-1$ and index $\frac12$, and since $\theta_{\frac12,\frac12}^1=i\eta^3$ (cf. (\ref{eqn:thetahalfhalf})) we may write
\begin{gather}\label{eqn:Z1P2varphieta6}
{Z^{(1)}_{\PP^2}(\tau,z)}={\overline{\varphi^{(1)}_{\PP^2}(\tau,\bar z)}}{\eta^{-6}(\tau)}
\end{gather}
where $\varphi^{(1)}_{\PP^2}(\tau,z)=t_{2,\frac12}(\tau,z)=\frac12t_{2,2}(\tau,\frac12z)$, and $t_{2,2}$ is the weight $2$, index $2$ skew-holomorphic Jacobi form that appears as a shadow in Mathieu moonshine (cf. (\ref{eqn:tmj})).

\medskip
The connection to Mathieu groups becomes stronger when we consider two M5-branes wrapping $\PP^2$. To explain this let 
$H(n)$ denote the Hurwitz class number of binary quadratic forms of discriminant $-n$ when $n>0$, and set $H(0):=-\frac1{12}$. 
Then $\mathscr{H}(\tau):=\sum_{n\geq 0}H(n)q^n$ is a mock modular form of weight $\frac32$ for $\Gamma_0(4)$ with shadow (proportional to) $\theta_{1,0}^0$ (cf. (\ref{eqn:thetamrj})). This was 
first discovered by Zagier \cite{Zag75}.
Very recent work \cite{CDM} proves that 
\begin{gather}\label{eqn:24H}
24\mathscr{H}(\tau)=-2+8q^3+12q^4+24q^7+24q^8+\dots
\end{gather}
is the graded dimension of a graded virtual module for the sporadic Mathieu group $M_{11}$, and $48\mathscr{H}(\tau)=-4+16q^3+24q^4+\dots$ is the graded dimension of a graded virtual module for $M_{23}$.
\medskip

Now set $\hat f_j(\tau):=3\hat h_j(\tau)\eta^{-6}(\tau)$ for $j\in \{0,1\}$, where $\hat h_j$ is the 
completion of the mock modular form 
\begin{gather}\label{eqn:hj}
h_j(\tau):=\sum_{n=0}^{\infty} H(4n+3j) q^{n + \frac{3j}4}.
\end{gather} 
Then the elliptic genus of two M5-branes wrapping $\PP^2$ is given \cite{Yoshioka94,Yoshioka95,VafaWitten} by 
$$Z_{{\mathbb P}^2}^{(2)}(\tau,z) = \hat f_0(\tau) 
\theta_{1,1}(-\bar\tau,-z)
- \hat f_1(\tau) 
\theta_{1,0}(-\bar\tau,-z).$$
Similar to (\ref{eqn:Z1P2varphieta6}) we may write
\begin{gather}\label{eqn:Z2P2varphieta6}
	Z^{(2)}_{\mathbb{P}^2}(\tau,z)={\overline{\varphi^{(2)}_{\PP^2}(\tau,\bar{z})}}{\eta^{-6}(\tau)}
\end{gather}
where $\varphi^{(2)}_{\PP^2}(\tau,z):=3\overline{\hat h_0(\tau)}\theta_{1,0}(\tau,z)-3\overline{\hat h_1(\tau)}\theta_{1,1}(\tau,z)$ is a skew-holomorphic mock Jacobi form of weight $2$ and index $1$ that exhibits moonshine for the Mathieu groups $M_{11}$ and $M_{23}$ according to \cite{CDM}.
Thus M5-branes on $\PP^2$ give a starting point from which we may pursue a geometric understanding of Mathieu moonshine for (rescaled) Hurwitz class numbers.

\medskip
It is interesting to note that the generating function $\mathscr{H}(\tau)$ 
also arises as an example of a 
function counting BPS jumping loci of maximal rank for $K3 \times T^2$, or equivalently, counting
attractor black holes, in the precise sense described in \cite{KT}. Also, the theta-coefficients of $\varphi$ in (\ref{eqn:Z2P2varphieta6}) recur in the elliptic genus for two M5-branes wrapping the Hirzebruch surface $\FF_1$ (see \S4.2 of \cite{Klemm}). 
In both these settings, and of course for two M5-branes wrapping $\PP^2$, it would be interesting to compare geometric twinings with the functions coming from the analysis of \cite{CDM}.

\subsection{Degree one del Pezzo surfaces}\label{sec:egs:dP8}

Next we consider M5-branes wrapping a del Pezzo surface 
of degree $1$ (i.e. $\PP^2$ blown up at $8$ points). The elliptic genus was first described in \cite{Yin}.
Start with the Fermat quintic
$\left\{\sum_i x_i^5 = 0\right\} \subset {\mathbb P}^4$
and quotient by the ${\mathbb Z}_5$ action
$x_i \to \omega^i x_i$ where $\omega := \ex(\frac15)$.
The hyperplane section ${\cal P}$ of the resulting orbifold 
has $\chi({\cal P})=11$ and is
rigid with $b_2^+ = 1$.  It has 
$H^2({\cal P},{\mathbb Z}) =  {\mathbb Z}\oplus (-E_8)$ 
and is thus a del Pezzo surface  of degree $1$.

\medskip
For 
a single M5-brane wrapping ${\cal P}$ we have $Z^{(1)}_{\dpe}(\tau,z)={\overline{\varphi^{(1)}_\dpe(\tau,\bar z)}}{\eta^{-6}(\tau)}$
for the elliptic genus,
where
\begin{gather}\label{eqn:varphiP}
\begin{split}
	\varphi^{(1)}_\dpe(\tau,z)
	:=&\,\overline{{E_4(\tau)}{\eta^{-8}(\tau)}}t_{2,\frac12}(\tau,z)\\
	=&\,\overline{{E_4(\tau)}{\eta^{-5}(\tau)}}
	(-i)\theta_{\frac12,\frac12}(\tau,z).
\end{split}
\end{gather}
This is a weakly skew-holomorphic Jacobi form of weight $2$ with a multiplier, and may be compared to (\ref{eqn:Z1P2varphieta6}). 

\medskip
Inspired by the discussion in \S\ref{sec:egs:P2} we consider the possibility that the coefficients of the anti-holomorphic factor in (\ref{eqn:varphiP}) also admit interpretations in terms of representations of Mathieu groups. Observe that 
\begin{gather}\label{eqn:f1cP}
f^{(1)}_\dpe(\tau):={E_4(\tau)}{\eta^{-5}(\tau)}
\end{gather}
is the unique modular form of weight $\frac32$ for $\SL_2(\ZZ)$ that has the same multiplier as $\eta^{-5}$ and satisfies $f^{(1)}_\dpe(\tau)=q^{-\frac5{24}}+O(q^{\frac{19}{24}})$ as $\Im(\tau)\to \infty$. By considering analogous functions for the congruence subgroups $\Gamma_0(n)<\SL_2(\ZZ)$ we are led to a family $f_{\dpe,nZ}^{(1)}$ 
of modular forms of weight $\frac32$ with various levels which achieves this goal for the sporadic simple Mathieu group $M_{12}$. That is, the $f^{(1)}_{\dpe,nZ}$ serve as trace functions 
\begin{gather}\label{eqn:WdP8}
	f^{(1)}_{\dpe,[g]}(\tau)=\sum_{d\in\ZZ+\frac{19}{24}}\tr\left(g|W^{(1)}_{\dpe,{d}}\right)q^{{d}}
\end{gather}
for a graded virtual $M_{12}$-module $W^{(1)}_{\dpe}=\bigoplus_{d} W^{(1)}_{\dpe,d}$
with graded dimension given by (\ref{eqn:f1cP}). 

\medskip
Details on the modular forms $f^{(1)}_{\dpe,nZ}$ are given in \S\ref{sec:numerics}, including the first few coefficients in their Fourier expansions (see Tables \ref{tab:coeffs:m12_1}-\ref{tab:coeffs:m12_2}) and the decompositions of the corresponding $W^{(1)}_{\dpe,{d}}$ into irreducible modules for $M_{12}$ (see Tables \ref{tab:mults:m12_1}-\ref{tab:mults:m12_5}). From that information alone it is not immediate that the virtual $M_{12}$-module $W^{(1)}_\dpe$ satisfying (\ref{eqn:WdP8}) exists, but we can verify this using arguments very similar to those appearing in recent literature on moonshine in weight $\frac32$, including \cite{DMO,CDM}. So we refrain from reproducing the details here.

\medskip
The reader will note that $f^{(1)}_\dpe$  is $\eta^3$ times the graded dimension of the basic representation $V_{E_8}$ of the affine Lie algebra of type $E_8$. This space naturally admits an action by the adjoint Lie group $E_8(\CC)$, so it is natural to ask if the twining functions $f^{(1)}_{\dpe,[g]}$ are related to this action. Here we note that $M_{12}$ is not a subgroup of $E_8(\CC)$ according to \cite{GR}, so the virtual $M_{12}$-module $W^{(1)}_\dpe$ cannot be recovered in a simple way from $V_{E_8}$. 

\medskip
We obtain an assignment of weakly skew-holomorphic Jacobi forms of weight $2$ and index $\frac12$ to elements of $M_{12}$ 
simply by setting 
\begin{gather}\label{eqn:varphi1PnZ}
\varphi^{(1)}_{\dpe,nZ}(\tau,z):=(-i)\overline{f^{(1)}_{\dpe,nZ}(\tau)}\theta_{\frac12,\frac12}(\tau,z).
\end{gather}
These forms in turn define twinings 
\begin{gather}\label{eqn:Z1PnZ}
Z^{(1)}_{\dpe,nZ}(\tau,z):=\overline{\varphi^{(1)}_{\dpe,nZ}(\tau,\bar{z})}\eta^{-6}(\tau)
\end{gather} 
of the M5-brane elliptic genus $Z^{(1)}_\dpe$.
As a result, it is natural to ask how the twining functions (\ref{eqn:Z1PnZ}) are related to the symmetries of M5-brane theory on $\cP$, and whether this relationship between $M_{12}$ and the del Pezzo surface of degree 1 is connected in some way to the original Mathieu moonshine \cite{EOT}, which relates $M_{24}$ to the K3 elliptic genus. It would be interesting to gain a physical or geometric understanding of the twining functions $Z^{(1)}_{\dpe,nZ}$.

\medskip
For a pair of M5-branes on $\cP$ we have
$Z^{(2)}_{\dpe}(\tau,z)={\overline{\varphi^{(2)}_\dpe(\tau,\bar z)}}{\eta^{-6}(\tau)}$ where
\begin{gather}\label{eqn:varphiPP}
	\varphi^{(2)}_\dpe(\tau,z):=	
\overline{{E_4^2(\tau)\eta^{-16}(\tau)}} 
\left(\overline{\hat h_0(\tau)}\theta_{1,0}(\tau,z)-\overline{\hat h_1(\tau)}\theta_{1,1}(\tau,z)\right)
\end{gather}
(cf. (\ref{eqn:hj})). In light of the discussions above and in \S\ref{sec:egs:P2} it seems likely that naturally defined Mathieu group twinings of $Z^{(2)}_\dpe$ also exist. Are there naturally defined twinings of $Z^{(n)}_\dpe$ by $g\in M_{12}$ for all $n$? What do they tell us about M5-branes on ${\cal P}$?

\subsection{Half-K3 surfaces}\label{sec:egs:halfK3}

In this final section we consider the elliptic genus for a single M5-brane wrapping a half-K3 surface
(i.e. ${\mathbb P}^2$ blown up at 9 points). Such surfaces play an important role in the study of E-strings via
geometric engineering. 

\medskip
To compute the genus in question we first discuss the cohomology group $H^2(\hkt,\ZZ)$. Geometrically it is generated by the hyperplane class corresponding to the hyperplane intersection with $\PP^2$, denoted by $H$, and the nine blow-ups $c_i$, for $i=1,\dots,9$. 
The quadratic form inherited from the intersection form is then given by ${\rm diag}(1,-1,\dots,-1)$. In fact, this lattice is isomorphic to $\ZZ\oplus(-\ZZ)\oplus (-E_8)$, and in particular is unimodular (but not even). 
The corresponding basis is given \cite{Estring} by 
\begin{gather}
\begin{split}
b_1 &:= 3H + \sum_{i=1}^8 c_i ~,~~b_2 :=- c_9,\\
e_8 &:= H+ \sum_{i=6}^8 c_i ~,~~e_i := c_i-c_{i+1}~{\rm{for}}~i=1,\dots 7.
\end{split}
\end{gather}

\medskip
Since the lattice is unimodular there is only one term in the decomposition (\ref{decomposetheta}) of the genus into theta functions. 
For the case at hand, the class ${\cal P}$ of the surface wrapped by the $M5$ brane is given by the anti-canonical class 
\be
{\cal P} = K_{\hkt} = b_1-b_2. 
\ee

\medskip
The theta function we are interested in will depend on the moduli of the half-K3, and one such modulus is given by the size of the elliptic fiber, denoted here by $\frac1R$. 
Taking the shift by $\frac12{\cal P}$ into account, the Siegel--Narain theta function (\ref{eqn:SieNartheta}) is $\Theta(R;\tau,z)=E_4(\tau)\Theta_{1,1}^{\rm odd}(R;\tau,z)$, where
\begin{gather}\label{eqn:halfK3ThetaRtauz}
\begin{split}
	&\Theta^{\rm odd}_{1,1}(R;\tau,z):=\\
	&\sum_{a,b\in\ZZ}(-1)^{a+b}
	q^{\frac1{2R^2}\left(R^2\frac{(a-b)}{2}+\frac{a+b+1}{2}\right)^2}
	\bar{q}^{\frac1{2R^2}\left(R^2\frac{(b-a)}{2}+\frac{a+b+1}{2}\right)^2}
	\bar{y}^{\left(R^2\frac{(b-a)}{2}+\frac{a+b+1}{2}\right)}.
\end{split}
\end{gather}
The elliptic genus is given by
\begin{gather}\label{eqn:Z1halfK3}
	Z_\hkt^{(1)}(R;\tau,z) = 
	{E_4(\tau)}
	\Theta^{\rm odd}_{1,1}(R;\tau,z){\eta^{-12}(\tau)}.
\end{gather}

\medskip
The question of moonshine-type phenomena is potentially richer in this setting due to the dependence on the parameter $R$. In this work we refrain from a full analysis and restrict ourselves to some special cases. In preparation for this note that (\ref{eqn:halfK3ThetaRtauz}) specializes to theta-type skew-holomorphic Jacobi forms of half-integral index (cf. (\ref{eqn:tmj})) at special values of $R$. Indeed, by a similar analysis to that given for $\Theta_{1,1}$ in \S\ref{sec:rationalGaussian} we obtain 
\begin{gather}\label{eqn:Thetaodd11sqrt2m}
\Theta^{\rm odd}_{1,1}(\sqrt{2m};\tau,z)=\overline{t_{1,m}(\tau,z)}
\end{gather}
when $m\in \ZZ+\frac12$ and $m>0$. 

\medskip
Motivated by the discussions in \S\ref{sec:egs:P2} and \S\ref{sec:egs:dP8} we now consider the decomposition 
${Z}^{(1)}_\hkt(\sqrt{2m};\tau)=\overline{\varphi^{(1)}_{\hkt,m}(\tau,z)}\eta^{-6}(\tau)$,
where by (\ref{eqn:Thetaodd11sqrt2m}) we have
\begin{gather}\label{eqn:varphi1hktm}
\varphi^{(1)}_{\hkt,m}(\tau,z)=\overline{E_4(\tau)\eta^{-6}(\tau)}t_{1,m}(\tau,z),
\end{gather}
which is a skew-holomorphic Jacobi form of weight $2$ and index $m$. 

\medskip
The first case to consider is $m=\frac12$, but $t_{1,\frac12}$ vanishes identically (cf. (\ref{eqn:thetahalfhalf})), so we set this case aside for the moment. The next case is $m=\frac32$, where, after applying 
(\ref{eqn:thetathreehalvespmhalf}) we find that 
\begin{gather}\label{eqn:varphi1hktthreehalves}
	\varphi^{(1)}_{\hkt,\frac32}(\tau,z)
	=\overline{E_4(\tau)\eta^{-5}(\tau)}(i)\left(\theta_{\frac32,\frac12}(\tau,z)-\theta_{\frac32,-\frac12}(\tau,z)\right).
\end{gather}
Observe that the anti-holomorphic factor in 
(\ref{eqn:varphi1hktthreehalves}) is precisely the same as that which appears in $\varphi^{(1)}_\dpe$ (cf. (\ref{eqn:varphiP})), in connection with del Pezzo surfaces of degree $1$. So from the discussion in \S\ref{sec:egs:dP8} we naturally obtain twinings 
\begin{gather}
	\varphi^{(1)}_{\hkt,\frac32,[g]}(\tau,z):=\overline{f^{(1)}_{\dpe,[g]}(\tau)}(i)\left(\theta_{\frac32,\frac12}(\tau,z)-\theta_{\frac32,-\frac12}(\tau,z)\right)
\end{gather}
(cf. (\ref{eqn:WdP8}))
of the weight $2$ skew-holomorphic Jacobi form (\ref{eqn:varphi1hktthreehalves}) 
by $g\in M_{12}$. This in turn leads to twinings 
\begin{gather}\label{eqn:tildeZ1halfK3g}
	{Z}^{(1)}_{\hkt,[g]}(\sqrt{3};\tau,z):=\overline{\varphi^{(1)}_{\hkt,\frac32,[g]}(\tau,z)}\eta^{-6}(\tau)
\end{gather}
by $g\in M_{12}$ of the single M5-brane elliptic genus for half-K3 surfaces at the modulus $R=\sqrt{3}$.

\medskip
The vanishing of (\ref{eqn:varphi1hktm}) 
at $m=\frac12$ suggests that we modify the elliptic genus by introducing a fermion number operator. This amounts to replacing $t_{1,m}$ with $t_{2,m}$ in (\ref{eqn:varphi1hktm}). Indeed, if we define
\begin{gather}\label{eqn:hatZ1hkt}
	\widetilde Z_\hkt^{(1)}(R;\tau,z) :=
	{E_4(\tau)}\widetilde\Theta_{1,1}^{\rm odd}(R;\tau,z){\eta^{-12}(\tau)}
\end{gather}
where 
\begin{gather}\label{eqn:halfK3hatThetaRtauz}
\begin{split}
	\widetilde\Theta_{1,1}^{\rm odd}(R;\tau,z)&:=
	\sum_{a,b\in\ZZ}
	(-1)^{a+b}
	\left(R^2\tfrac{(a-b)}{2}+\tfrac{a+b+1}{2}\right)
	q^{\frac1{2R^2}\left(R^2\frac{(a-b)}{2}+\frac{a+b+1}{2}\right)^2}\\
	&\qquad\qquad\qquad\times
	\bar{q}^{\frac1{2R^2}\left(R^2\frac{(b-a)}{2}+\frac{a+b+1}{2}\right)^2}
	\bar{y}^{\left(R^2\frac{(b-a)}{2}+\frac{a+b+1}{2}\right)}
\end{split}
\end{gather}
then
$\widetilde\Theta_{1,1}^{\rm odd}(\sqrt{2m};\tau,z)=\overline{t_{2,m}(\tau,z)}$
when $m\in \ZZ+\frac12$ and $m>0$. In this setting we consider the decomposition
$\widetilde{Z}^{(1)}_\hkt(\sqrt{2m};\tau)=\overline{\widetilde\varphi^{(1)}_{\hkt,m}(\tau,z)}\eta^{-4}(\tau)$
so that
\begin{gather}\label{eqn:hatvarphi1hktm}
\widetilde\varphi^{(1)}_{\hkt,m}(\tau,z)=\overline{E_4(\tau)\eta^{-8}(\tau)}t_{2,m}(\tau,z)
\end{gather}
is the associated weakly skew-holomorphic Jacobi form of weight $2$ and index $m$.

\medskip
Now comparing with (\ref{eqn:varphiP}) we find that (\ref{eqn:hatvarphi1hktm}) 
at $m=\frac12$
is precisely the skew-holomorphic Jacobi form of weight $2$ and index $\frac12$ 
that appeared in \S\ref{sec:egs:dP8} in connection with del Pezzo surfaces of degree $1$. So it is natural to define 
\begin{gather}
\widetilde\varphi^{(1)}_{\hkt,\frac12,[g]}(\tau,z):=\varphi^{(1)}_{\dpe,[g]}(\tau,z)
\end{gather} for $g\in M_{12}$. We then obtain twinings 
\begin{gather}\label{eqn:hatZ1halfK3g}
\widetilde{Z}^{(1)}_{\hkt,[g]}(1;\tau):=\overline{\widetilde\varphi^{(1)}_{\hkt,\frac12,[g]}(\tau,z)}\eta^{-4}(\tau)
\end{gather}
of the modified single M5-brane elliptic genus for half-K3 surfaces by $g\in M_{12}$ when $R=1$.

\medskip
We conclude this section with four remarks. 
Firstly, 
it is natural to ask if twinings by $M_{12}$ of 
the elliptic genera (\ref{eqn:Z1halfK3}) and (\ref{eqn:hatZ1hkt}) for half-K3 surfaces can be defined for all $R$. 
Are there special values of $R$ for which this hidden symmetry extends beyond $M_{12}$?
Secondly, 
it would be interesting to compare the physical twinings of the elliptic genera  (\ref{eqn:Z1halfK3}) and (\ref{eqn:hatZ1hkt}) with the series (\ref{eqn:tildeZ1halfK3g}) and (\ref{eqn:hatZ1halfK3g}) that arise from $M_{12}$ in the manner we have just described. 
Thirdly, note that the given expressions (\ref{eqn:Z1halfK3}) and (\ref{eqn:hatZ1hkt}) for the elliptic genera considered in this section point to explicit realizations in terms of the vertex algebra attached to the lattice $\ZZ\oplus (-\ZZ)\oplus (-E_8)$. It would be interesting to determine if the twinings (\ref{eqn:tildeZ1halfK3g}) and (\ref{eqn:hatZ1halfK3g}) by elements of $M_{12}$ can also be realized using this structure.
In view of the well-known Mathieu moonshine connection between $M_{24}$ and K3 surfaces \cite{EOT}, it is appealing that the Euler characteristic of a half-K3 surface is $12$. So finally we ask, to what extent is the connection between half-K3 surfaces and $M_{12}$ described in this section related to the original Mathieu moonshine?

\bigskip
\centerline{\bf{Acknowledgements}}
\medskip
We are grateful to A. Klemm and J. Manschot for useful correspondence, and to G. Moore for helpful discussions and comments on an earlier draft.
We would like to thank the American Institute of Mathematics for hosting four of us (MC, JD, SH and SK)  as members of a SQuaRE working group on ``Moonshine and String Theory" in May of 2017. Four of us (MC, JH, SH, and SK) also acknowledge the kind hospitality of the Aspen Center for Physics, which is supported by NSF grant PHY-1066293, as this was being completed. 
The work of M.C. was supported by ERC starting grant H2020 ERC StG \#640159. 
J.D. acknowledges support from the U.S. National Science Foundation (DMS 1601306), and the Simons Foundation (\#316779). 
J.H. acknowledges support from the U.S. National Science Foundation (PHY 1520748) and from the Simons Foundation (\#399639).
S.H. is supported by a Harvard University Golub Fellowship in the Physical Sciences and DOE grant DE-SC0007870.
S.K. is partially supported by a Simons Investigator Award and by the U.S. National Science Foundation under grant PHY-1720397.

\appendix
\section{New Mathieu moonshine}\label{sec:numerics}

Here we present numerical data in support of the discussions of \S\ref{sec:egs:dP8} and \S\ref{sec:egs:halfK3} relating the M5-brane elliptic genera for degree one del Pezzo surfaces and half-K3 surfaces to the Mathieu group $M_{12}$. Since the relationship to half-K3 surfaces  is formulated in terms of the functions that appear in \S\ref{sec:egs:dP8} we employ the notation of \S\ref{sec:egs:dP8} in what follows. 

\medskip
The character table of $M_{12}$ is Table \ref{tab:chars:M12}, wherein $b_{11}:=-\frac12+\frac{\sqrt{-11}}2$. Tables \ref{tab:coeffs:m12_1}-\ref{tab:coeffs:m12_2} give the coefficients of $q^d$ in the McKay--Thompson series $f^{(1)}_{\dpe,nZ}$ up to $d=\frac{2275}{24}$. The naming of the conjugacy classes is as in Table \ref{tab:chars:M12}. Tables \ref{tab:mults:m12_1}-\ref{tab:mults:m12_5} give the multiplicity generating functions for irreducible characters in the graded virtual $M_{12}$-module $W^{(1)}_{\dpe,d}$ of (\ref{eqn:WdP8}). 
That is, for $\chi$ an irreducible character of $M_{12}$, the coefficient of $q^{d}$ in $f^{(1)}_{\dpe,\chi}$ denotes the multiplicity of $\chi$ in the (virtual) $M_{12}$-module $W^{(1)}_{\dpe,d}$. In Tables \ref{tab:mults:m12_1}-\ref{tab:mults:m12_5} the characters are named by their dimensions, and appear in the same order as in Table \ref{tab:chars:M12}.

\medskip
The modular forms $f^{(1)}_{\dpe,nZ}$ may be realized as Rademacher sums. Specifically, consider the degree $24$ permutation representation of $M_{12}$ that arises by restricting the defining permutation representation of $M_{24}$. The corresponding character is $2\chi_1+\chi_2+\chi_3$ in the notation of Table \ref{tab:chars:M12}. If $g\in M_{12}$ has order $n$ and $h$ is the minimal length of a cycle in the cycle shape of $g$ (in this permutation representation) then $f^{(1)}_{\dpe,[g]}$ is the Rademacher sum of weight $\frac32$ for $\Gamma_0(n)$ with polar part $q^{-\frac{5}{24}}$ (at the infinite cusp), and multiplier system given by $\gamma\mapsto \ex(\frac{cd}{nh})\epsilon^{-5}(\gamma)$ for $\gamma=\left(\begin{smallmatrix} *&*\\c&d\end{smallmatrix}\right)\in\Gamma_0(n)$, where $\epsilon$ is the multiplier system of $\eta$. The values of $n$ and $h$ for each conjugacy class $[g]\subset M_{12}$ are given in Table \ref{tab:chars:M12}. We refer to \cite{ChengDuncanRevRad} for details of the Rademacher sum construction. 

\medskip
Note that since the factor $\ex(\frac{cd}{nh})$ is trivial on $\Gamma_0(nh)$ the function $f^{(1)}_{\dpe,[g]}\eta^5$ is a holomorphic modular form of weight $4$ for $\Gamma_0(nh)$. This, together with the Fourier coefficients in Tables \ref{tab:coeffs:m12_1}-\ref{tab:coeffs:m12_2} gives an alternative method for reconstructing the $f^{(1)}_{\dpe,nZ}$. For example, we find in this way that 
\begin{gather}
\begin{split}
f^{(1)}_{\dpe,\mathrm{2B}}(\tau)&=\tfrac{1}{15}(16E_4(2\tau)-E_4(\tau))\eta^{-5}(\tau),\\
f^{(1)}_{\dpe,\mathrm{3A}}(\tau)&=\tfrac{1}{80}(81E_4(3\tau)-E_4(\tau))\eta^{-5}(\tau),\\
f^{(1)}_{\dpe,\mathrm{2A}}(\tau)&=f^{(1)}_{\dpe,2B}(\tau)+32\eta^3(\tau)\eta(4\tau)^8\eta^{-8}(2\tau).
\end{split}
\end{gather}

\begin{table}[ht]
\begin{center}
\caption{\small{Character table of $M_{12}$}}\label{tab:chars:M12}
\begin{small}
\medskip
\begin{tabular}{c@{\;\;}|r@{\;\;}r@{\;\;}r@{\;\;}r@{\;\;}r@{\;\;}r@{\;\;}r@{\;\;}r@{\;\;}r@{\;\;}r@{\;\;}r@{\;\;}r@{\;\;}r@{\;\;}r@{\;\;}r}
\toprule
$[g]$&   1A&   2A&   2B&   3A&   3B&   4A&   4B&   5A&   6A&   6B&   8A&   8B&   10A&   11A&   11B\\
\midrule
$n|h$&   $1|1$&   $2|2$&   $2|1$&   $3|1$&   $3|3$&   $4|1$&   $4|1$&   $5|1$&   $6|6$&   $6|1$&   $8|1$&   $8|1$&   $10|2$&   $11|1$&   $11|1$\\
\midrule
${\chi}_{1}$&   $1$&   $1$&   $1$&   $1$&   $1$&   $1$&   $1$&   $1$&   $
1$&   $1$&   $1$&   $1$&   $1$&   $1$&   $1$\\
${\chi}_{2}$&   $11$&   $-1$&   $3$&   $2$&   $-1$&   $-1$&   $3$&   $1$&   $
-1$&   $0$&   $-1$&   $1$&   $-1$&   $0$&   $0$\\
${\chi}_{3}$&   $11$&   $-1$&   $3$&   $2$&   $-1$&   $3$&   $-1$&   $1$&   $
-1$&   $0$&   $1$&   $-1$&   $-1$&   $0$&   $0$\\
${\chi}_{4}$&   $16$&   $4$&   $0$&   $-2$&   $1$&   $0$&   $0$&   $1$&   $
1$&   $0$&   $0$&   $0$&   $-1$&   $b_{11}$&   $\overline{b_{11}}$\\
${\chi}_{5}$&   $16$&   $4$&   $0$&   $-2$&   $1$&   $0$&   $0$&   $1$&   $
1$&   $0$&   $0$&   $0$&   $-1$&   $\overline{b_{11}}$&   $b_{11}$\\
${\chi}_{6}$&   $45$&   $5$&   $-3$&   $0$&   $3$&   $1$&   $1$&   $0$&   $
-1$&   $0$&   $-1$&   $-1$&   $0$&   $1$&   $1$\\
${\chi}_{7}$&   $54$&   $6$&   $6$&   $0$&   $0$&   $2$&   $2$&   $-1$&   $
0$&   $0$&   $0$&   $0$&   $1$&   $-1$&   $-1$\\
${\chi}_{8}$&   $55$&   $-5$&   $7$&   $1$&   $1$&   $-1$&   $-1$&   $0$&   $
1$&   $1$&   $-1$&   $-1$&   $0$&   $0$&   $0$\\
${\chi}_{9}$&   $55$&   $-5$&   $-1$&   $1$&   $1$&   $3$&   $-1$&   $0$&   $
1$&   $-1$&   $-1$&   $1$&   $0$&   $0$&   $0$\\
${\chi}_{10}$&   $55$&   $-5$&   $-1$&   $1$&   $1$&   $-1$&   $3$&   $
0$&   $1$&   $-1$&   $1$&   $-1$&   $0$&   $0$&   $0$\\
${\chi}_{11}$&   $66$&   $6$&   $2$&   $3$&   $0$&   $-2$&   $-2$&   $1$&   $
0$&   $-1$&   $0$&   $0$&   $1$&   $0$&   $0$\\
${\chi}_{12}$&   $99$&   $-1$&   $3$&   $0$&   $3$&   $-1$&   $-1$&   $
-1$&   $-1$&   $0$&   $1$&   $1$&   $-1$&   $0$&   $0$\\
${\chi}_{13}$&   $120$&   $0$&   $-8$&   $3$&   $0$&   $0$&   $0$&   $0$&   $
0$&   $1$&   $0$&   $0$&   $0$&   $-1$&   $-1$\\
${\chi}_{14}$&   $144$&   $4$&   $0$&   $0$&   $-3$&   $0$&   $0$&   $
-1$&   $1$&   $0$&   $0$&   $0$&   $-1$&   $1$&   $1$\\
${\chi}_{15}$&   $176$&   $-4$&   $0$&   $-4$&   $-1$&   $0$&   $0$&   $
1$&   $-1$&   $0$&   $0$&   $0$&   $1$&   $0$&   $0$\\
\bottomrule
\end{tabular}
\end{small}
\end{center}
\end{table}

\clearpage

\begin{sidewaystable}[ht]
\vspace{-8pt}
\begin{scriptsize}
\begin{center}
\caption{\small{McKay--Thompson series $f^{(1)}_{\dpe,nZ}$, part 1}}\label{tab:coeffs:m12_1}\medskip
\begin{tabular}{c|@{ }r@{ }r@{ }r@{ }r@{ }r@{ }r@{ }r@{ }r@{ }r@{ }r@{ }r@{ }r}\toprule
$24d$	&1A	&2A	&2B	&3A	&3B	&4AB	&5A	&6A	&6B	&8AB	&10A	&11AB	\\
\midrule
-5	&$1$	&$1$	&$1$	&$1$	&$1$	&$1$	&$1$	&$1$	&$1$	&$1$	&$1$	&$1$	\\
19	&$245$	&$21$	&$-11$	&$2$	&$-7$	&$5$	&$-5$	&$-3$	&$-2$	&$1$	&$1$	&$3$	\\
43	&$3380$	&$-44$	&$52$	&$-22$	&$14$	&$4$	&$5$	&$-2$	&$-2$	&$0$	&$1$	&$-8$	\\
67	&$22385$	&$113$	&$-143$	&$29$	&$11$	&$-15$	&$10$	&$11$	&$1$	&$1$	&$-2$	&$0$	\\
91	&$110110$	&$-322$	&$286$	&$31$	&$-77$	&$-18$	&$-15$	&$-1$	&$7$	&$-6$	&$-7$	&$0$	\\
115	&$438746$	&$602$	&$-550$	&$-112$	&$77$	&$26$	&$-4$	&$-7$	&$8$	&$-6$	&$2$	&$0$	\\
139	&$1531985$	&$-1071$	&$1105$	&$113$	&$77$	&$17$	&$-15$	&$-3$	&$-11$	&$1$	&$-1$	&$4$	\\
163	&$4804910$	&$1870$	&$-2002$	&$71$	&$-253$	&$-66$	&$35$	&$-17$	&$-13$	&$-6$	&$-5$	&$0$	\\
187	&$13914285$	&$-3283$	&$3245$	&$-381$	&$231$	&$-19$	&$35$	&$11$	&$-13$	&$13$	&$7$	&$11$	\\
211	&$37674325$	&$5525$	&$-5291$	&$334$	&$145$	&$117$	&$-50$	&$29$	&$10$	&$13$	&$0$	&$7$	\\
235	&$96580627$	&$-8621$	&$8723$	&$277$	&$-704$	&$51$	&$2$	&$16$	&$29$	&$-5$	&$4$	&$0$	\\
259	&$236144545$	&$13377$	&$-13663$	&$-911$	&$637$	&$-143$	&$-80$	&$-15$	&$17$	&$13$	&$12$	&$-23$	\\
283	&$554578570$	&$-20790$	&$20618$	&$811$	&$469$	&$-86$	&$70$	&$-39$	&$-1$	&$-14$	&$10$	&$-4$	\\
307	&$1256789730$	&$31458$	&$-31006$	&$534$	&$-1617$	&$226$	&$105$	&$-21$	&$-46$	&$-14$	&$-7$	&$-2$	\\
331	&$2760379655$	&$-46073$	&$46343$	&$-2173$	&$1364$	&$135$	&$-95$	&$4$	&$-25$	&$7$	&$-13$	&$0$	\\
355	&$5894771883$	&$67179$	&$-67925$	&$1824$	&$924$	&$-373$	&$8$	&$48$	&$16$	&$-13$	&$4$	&$0$	\\
379	&$12275038600$	&$-97752$	&$97416$	&$1249$	&$-3575$	&$-168$	&$-150$	&$33$	&$45$	&$20$	&$-12$	&$0$	\\
403	&$24982062560$	&$139584$	&$-138528$	&$-4420$	&$2915$	&$528$	&$185$	&$-9$	&$36$	&$20$	&$-11$	&$-7$	\\
427	&$49794727675$	&$-196133$	&$196603$	&$3595$	&$2002$	&$235$	&$175$	&$-62$	&$-29$	&$-1$	&$7$	&$0$	\\
451	&$97369902630$	&$274022$	&$-275418$	&$2376$	&$-6930$	&$-698$	&$-245$	&$-46$	&$-72$	&$14$	&$7$	&$11$	\\
475	&$187076653120$	&$-381024$	&$380224$	&$-8738$	&$5698$	&$-400$	&$-5$	&$42$	&$-62$	&$-44$	&$1$	&$38$	\\
499	&$353616436085$	&$524277$	&$-522379$	&$7079$	&$3641$	&$949$	&$-290$	&$81$	&$35$	&$-43$	&$12$	&$0$	\\
523	&$658376681690$	&$-713734$	&$714714$	&$4727$	&$-13300$	&$490$	&$315$	&$80$	&$123$	&$6$	&$11$	&$-31$	\\
547	&$1208616966765$	&$966093$	&$-968851$	&$-16155$	&$10593$	&$-1379$	&$390$	&$-51$	&$89$	&$-31$	&$-22$	&$32$	\\
571	&$2189664565985$	&$-1302879$	&$1301729$	&$12908$	&$7004$	&$-575$	&$-390$	&$-144$	&$-40$	&$57$	&$-24$	&$-27$	\\
595	&$3918286118747$	&$1744155$	&$-1740453$	&$8147$	&$-23947$	&$1851$	&$-3$	&$-111$	&$-141$	&$51$	&$5$	&$0$	\\
619	&$6930554664880$	&$-2316048$	&$2317744$	&$-29150$	&$18865$	&$848$	&$-495$	&$45$	&$-98$	&$-24$	&$-13$	&$0$	\\
643	&$12125024699095$	&$3062199$	&$-3066921$	&$22825$	&$11935$	&$-2361$	&$595$	&$183$	&$45$	&$43$	&$-21$	&$0$	\\
667	&$20994476982895$	&$-4034001$	&$4031599$	&$14815$	&$-42092$	&$-1201$	&$645$	&$132$	&$175$	&$-57$	&$29$	&$-24$	\\
691	&$35997712990855$	&$5283495$	&$-5277305$	&$-50087$	&$32956$	&$3095$	&$-770$	&$-60$	&$133$	&$-45$	&$20$	&$0$	\\
715	&$61152257741861$	&$-6880923$	&$6883877$	&$39062$	&$21098$	&$1477$	&$-14$	&$-234$	&$-106$	&$13$	&$2$	&$11$	\\
739	&$102971911295570$	&$8927122$	&$-8935342$	&$24530$	&$-71194$	&$-4110$	&$-930$	&$-134$	&$-250$	&$-38$	&$32$	&$42$	\\
763	&$171940936021855$	&$-11542209$	&$11538527$	&$-84881$	&$55132$	&$-1841$	&$980$	&$120$	&$-193$	&$83$	&$36$	&$0$	\\
787	&$284816074366495$	&$14858431$	&$-14847713$	&$65539$	&$34552$	&$5359$	&$1120$	&$280$	&$127$	&$75$	&$-24$	&$-54$	\\
811	&$468201092136435$	&$-19041645$	&$19046643$	&$41319$	&$-118503$	&$2499$	&$-1190$	&$213$	&$339$	&$-1$	&$-20$	&$31$	\\
835	&$764062908896885$	&$24320821$	&$-24334219$	&$-138991$	&$90923$	&$-6699$	&$10$	&$-125$	&$233$	&$45$	&$-4$	&$-51$	\\
859	&$1238199118586430$	&$-30972578$	&$30966078$	&$106548$	&$57057$	&$-3250$	&$-1445$	&$-407$	&$-96$	&$-118$	&$-43$	&$0$	\\
883	&$1993162922073180$	&$39301308$	&$-39284388$	&$66213$	&$-191268$	&$8460$	&$1680$	&$-264$	&$-423$	&$-112$	&$-32$	&$0$	\\
907	&$3187894582604875$	&$-49688021$	&$49696075$	&$-224249$	&$146146$	&$4027$	&$1750$	&$118$	&$-269$	&$31$	&$34$	&$0$	\\
931	&$5067385763330905$	&$62640249$	&$-62662055$	&$170881$	&$90349$	&$-10903$	&$-1970$	&$477$	&$157$	&$-75$	&$44$	&$-8$	\\
955	&$8007296783387517$	&$-78760899$	&$78751101$	&$106299$	&$-304920$	&$-4899$	&$17$	&$324$	&$507$	&$149$	&$1$	&$0$	\\
979	&$12580745731759205$	&$98730149$	&$-98702747$	&$-353800$	&$231308$	&$13701$	&$-2420$	&$-196$	&$352$	&$125$	&$44$	&$55$	\\
1003	&$19657853187268125$	&$-123390883$	&$123403293$	&$268005$	&$143418$	&$6205$	&$2500$	&$-574$	&$-243$	&$-43$	&$32$	&$93$	\\
1027	&$30553454673736995$	&$153823299$	&$-153856989$	&$164577$	&$-475629$	&$-16845$	&$2870$	&$-429$	&$-639$	&$111$	&$-46$	&$0$	\\
1051	&$47245393529211705$	&$-191308807$	&$191292729$	&$-550938$	&$359007$	&$-8039$	&$-3045$	&$299$	&$-462$	&$-175$	&$-67$	&$-96$	\\
1075	&$72695798459621870$	&$237317774$	&$-237275922$	&$414887$	&$219776$	&$20926$	&$-5$	&$752$	&$267$	&$-150$	&$-1$	&$38$	\\
1099	&$111322113952614145$	&$-293640095$	&$293659649$	&$255274$	&$-732809$	&$9777$	&$-3605$	&$559$	&$818$	&$37$	&$-45$	&$-57$	\\
1123	&$169685042685799025$	&$362516273$	&$-362568591$	&$-842182$	&$550550$	&$-26159$	&$4025$	&$-310$	&$558$	&$-119$	&$-67$	&$0$	\\\bottomrule
\end{tabular}
\end{center}
\end{scriptsize}
\end{sidewaystable}

\clearpage

\begin{sidewaystable}[ht]
\vspace{-8pt}
\begin{scriptsize}
\begin{center}
\caption{\small{McKay--Thompson series $f^{(1)}_{\dpe,nZ}$, part 2}}\label{tab:coeffs:m12_2}\medskip
\begin{tabular}{c|@{ }r@{ }r@{ }r@{ }r@{ }r@{ }r@{ }r@{ }r@{ }r@{ }r@{ }r@{ }r}\toprule
$24d$	&1A	&2A	&2B	&3A	&3B	&4AB	&5A	&6A	&6B	&8AB	&10A	&11AB	\\
\midrule
1147	&$257489647589909735$	&$-446610297$	&$446586855$	&$631943$	&$337337$	&$-11721$	&$4235$	&$-963$	&$-273$	&$195$	&$63$	&$0$	\\
1171	&$389036935666670005$	&$548983797$	&$-548919371$	&$384994$	&$-1110824$	&$32213$	&$-4745$	&$-648$	&$-1010$	&$189$	&$57$	&$0$	\\
1195	&$585321427870953137$	&$-673325583$	&$673354929$	&$-1274347$	&$830753$	&$14673$	&$12$	&$333$	&$-699$	&$-39$	&$-8$	&$1$	\\
1219	&$877051946642599585$	&$824194401$	&$-824272735$	&$951751$	&$505153$	&$-39167$	&$-5540$	&$1125$	&$383$	&$121$	&$56$	&$0$	\\
1243	&$1308987868412927175$	&$-1006964089$	&$1006927559$	&$579426$	&$-1666434$	&$-18265$	&$5925$	&$770$	&$1178$	&$-273$	&$81$	&$66$	\\
1267	&$1946143827213299420$	&$1227840124$	&$-1227744804$	&$-1900084$	&$1241051$	&$47660$	&$6545$	&$-413$	&$852$	&$-248$	&$-71$	&$80$	\\
1291	&$2882640700620305625$	&$-1494261447$	&$1494305241$	&$1413606$	&$753753$	&$21897$	&$-6875$	&$-1347$	&$-534$	&$45$	&$-57$	&$0$	\\
1315	&$4254293388075663716$	&$1815256100$	&$-1815372956$	&$855485$	&$-2465848$	&$-58428$	&$-34$	&$-964$	&$-1451$	&$-180$	&$0$	&$-151$	\\
1339	&$6256460810550333175$	&$-2201441577$	&$2201389047$	&$-2806274$	&$1830877$	&$-26265$	&$-8075$	&$573$	&$-1026$	&$347$	&$-87$	&$34$	\\
1363	&$9169285487540907855$	&$2665121487$	&$-2664979633$	&$2080578$	&$1105104$	&$70927$	&$8855$	&$1656$	&$602$	&$303$	&$-73$	&$-105$	\\
1387	&$13393282867503977295$	&$-3220900017$	&$3220964175$	&$1257615$	&$-3616767$	&$32079$	&$9420$	&$1173$	&$1839$	&$-97$	&$68$	&$0$	\\
1411	&$19499387894072616035$	&$3886321059$	&$-3886491037$	&$-4096807$	&$2675288$	&$-84989$	&$-10340$	&$-708$	&$1265$	&$235$	&$104$	&$0$	\\
1435	&$28299145776890952502$	&$-4681963594$	&$4681885494$	&$3027835$	&$1613458$	&$-39050$	&$2$	&$-2050$	&$-645$	&$-362$	&$6$	&$0$	\\
1459	&$40942900182985321885$	&$5631643773$	&$-5631439203$	&$1820005$	&$-5246318$	&$102285$	&$-11990$	&$-1398$	&$-2151$	&$-319$	&$88$	&$-32$	\\
1483	&$59056800250327174295$	&$-6763471561$	&$6763564951$	&$-5930560$	&$3868865$	&$46695$	&$12670$	&$713$	&$-1460$	&$99$	&$94$	&$0$	\\
1507	&$84933501203533915790$	&$8110930190$	&$-8111177074$	&$4368671$	&$2321594$	&$-123442$	&$13790$	&$2474$	&$791$	&$-242$	&$-110$	&$121$	\\
1531	&$121796962409191079920$	&$-9713117456$	&$9713006576$	&$2623747$	&$-7548422$	&$-55440$	&$-14705$	&$1666$	&$2519$	&$448$	&$-111$	&$184$	\\
1555	&$174169266048720253146$	&$11615262618$	&$-11614967078$	&$-8500497$	&$5550897$	&$147770$	&$21$	&$-891$	&$1759$	&$386$	&$-7$	&$0$	\\
1579	&$248377592707632513150$	&$-13870493250$	&$13870626430$	&$6245646$	&$3326730$	&$66590$	&$-16850$	&$-2850$	&$-1094$	&$-58$	&$-100$	&$-206$	\\
1603	&$353253324430055413910$	&$16541557942$	&$-16541909098$	&$3734813$	&$-10760596$	&$-175578$	&$18410$	&$-1988$	&$-3067$	&$274$	&$-118$	&$106$	\\
1627	&$501093959691581866120$	&$-19701491768$	&$19701331592$	&$-12095267$	&$7891807$	&$-80088$	&$19495$	&$1231$	&$-2191$	&$-576$	&$147$	&$-160$	\\
1651	&$708983776598335157920$	&$23434722304$	&$-23434304608$	&$8862610$	&$4711630$	&$208848$	&$-21080$	&$3418$	&$1250$	&$-524$	&$124$	&$0$	\\
1675	&$1000603206531275493740$	&$-27839878548$	&$27840067948$	&$5292041$	&$-15230776$	&$94700$	&$-10$	&$2424$	&$3709$	&$124$	&$2$	&$0$	\\
1699	&$1408702624118535140395$	&$33032707435$	&$-33033205205$	&$-17064938$	&$11141053$	&$-248885$	&$-24230$	&$-1379$	&$2530$	&$-365$	&$160$	&$0$	\\
1723	&$1978477667607266537870$	&$-39147551762$	&$39147329166$	&$12473543$	&$6641696$	&$-111298$	&$25620$	&$-4172$	&$-1317$	&$674$	&$148$	&$-7$	\\
1747	&$2772165465354178800390$	&$46339376966$	&$-46338787066$	&$7423860$	&$-21384528$	&$294950$	&$27765$	&$-2872$	&$-4360$	&$606$	&$-139$	&$0$	\\
1771	&$3875291180776298187905$	&$-54788436927$	&$54788701825$	&$-23916919$	&$15607823$	&$132449$	&$-29345$	&$1491$	&$-2951$	&$-183$	&$-137$	&$121$	\\
1795	&$5405141185164404628126$	&$64705171230$	&$-64705866082$	&$17441601$	&$9274839$	&$-347426$	&$1$	&$4887$	&$1649$	&$462$	&$5$	&$226$	\\
1819	&$7522234988258726535800$	&$-76332974536$	&$76332659832$	&$10365428$	&$-29834728$	&$-157352$	&$-33575$	&$3284$	&$5076$	&$-768$	&$-191$	&$0$	\\

1843	&$10445828606352623315780$	&$89952081860$	&$-89951262908$	&$-33278401$	&$21724703$	&$409476$	&$36155$	&$-1885$	&$3559$	&$-652$	&$-165$	&$-261$	\\
1867	&$14474828224085044461275$	&$-105887271301$	&$105887640027$	&$24216572$	&$12892502$	&$184363$	&$38150$	&$-5626$	&$-2112$	&$167$	&$154$	&$124$	\\
1891	&$20015952136902352017085$	&$124515746461$	&$-124516712515$	&$14350090$	&$-41334620$	&$-483027$	&$-41165$	&$-3956$	&$-5998$	&$-495$	&$191$	&$-203$	\\
1915	&$27621587036170435738716$	&$-146272523812$	&$146272092252$	&$-46027323$	&$30036996$	&$-215780$	&$-34$	&$2336$	&$-4179$	&$892$	&$-12$	&$0$	\\
1939	&$38040588686504509649390$	&$171657529358$	&$-171656394514$	&$33423194$	&$17775758$	&$567422$	&$-46735$	&$6710$	&$2378$	&$810$	&$173$	&$0$	\\
1963	&$52286338577292264775330$	&$-201247917374$	&$201248425378$	&$19779820$	&$-56935802$	&$254002$	&$49455$	&$4666$	&$7156$	&$-162$	&$171$	&$0$	\\
1987	&$71727767470164471985970$	&$235711210322$	&$-235712537038$	&$-63253216$	&$41292251$	&$-663358$	&$53095$	&$-2629$	&$4868$	&$558$	&$-213$	&$-75$	\\
2011	&$98210898877425806480580$	&$-275815152540$	&$275814555588$	&$45846324$	&$24403500$	&$-298476$	&$-56045$	&$-7932$	&$-2592$	&$-1104$	&$-215$	&$0$	\\
2035	&$134220886793221405680657$	&$322440005713$	&$-322438454767$	&$27064422$	&$-77947254$	&$775473$	&$32$	&$-5402$	&$-8362$	&$-999$	&$-12$	&$187$	\\
2059	&$183097700277061507239685$	&$-376598728955$	&$376599424773$	&$-86460161$	&$56425369$	&$347909$	&$-63565$	&$2905$	&$-5649$	&$229$	&$-205$	&$328$	\\
2083	&$249322773322218559282605$	&$439458518285$	&$-439460332883$	&$62548884$	&$33271029$	&$-907299$	&$67980$	&$9221$	&$3160$	&$-703$	&$-240$	&$0$	\\
2107	&$338899391506589069707460$	&$-512358042588$	&$512357234116$	&$36873998$	&$-106155721$	&$-404236$	&$71960$	&$6279$	&$9562$	&$1288$	&$252$	&$-399$	\\
2131	&$459856714942619762629680$	&$596828898736$	&$-596826783696$	&$-117510219$	&$76706784$	&$1057520$	&$-76945$	&$-3464$	&$6585$	&$1104$	&$241$	&$203$	\\
2155	&$622916642534098871009415$	&$-694628315129$	&$694629258375$	&$84864729$	&$45168123$	&$471623$	&$40$	&$-10613$	&$-3831$	&$-313$	&$-4$	&$-329$	\\
2179	&$842374854253122861597685$	&$807773972437$	&$-807776428811$	&$49926859$	&$-143781275$	&$-1228187$	&$-86815$	&$-7307$	&$-11201$	&$825$	&$247$	&$0$	\\
2203	&$1137263164627672249293665$	&$-938574065919$	&$938572965473$	&$-158926840$	&$103724390$	&$-550223$	&$91665$	&$4242$	&$-7780$	&$-1427$	&$301$	&$0$	\\
2227	&$1532880862104682603160505$	&$1089663943033$	&$-1089661090375$	&$114584796$	&$60954696$	&$1426329$	&$97755$	&$12388$	&$4364$	&$-1247$	&$-257$	&$0$	\\
2251	&$2062809388122724224605390$	&$-1264058425906$	&$1264059700430$	&$67322939$	&$-193818955$	&$637262$	&$-103235$	&$8561$	&$13199$	&$302$	&$-261$	&$-67$	\\
2275	&$2771559316831560463943735$	&$1465208427543$	&$-1465211739849$	&$-213852964$	&$139592453$	&$-1656153$	&$-15$	&$-4851$	&$9012$	&$-901$	&$-7$	&$0$	\\  \bottomrule
\end{tabular}
\end{center}
\end{scriptsize}
\end{sidewaystable}

\clearpage

\begin{sidewaystable}[ht]
\vspace{-8pt}
\begin{scriptsize}
\begin{center}
\caption{\small{Multiplicity generating functions $f^{(1)}_{\dpe,\chi}$, part 1}}\label{tab:mults:m12_1}\medskip
\begin{tabular}{c|@{ }r@{ }r@{ }r@{ }r@{ }r@{ }r@{ }r@{ }r}\toprule
$24d$	&$1$	&$11$	&$11'$	&$16$	&${16'}$	&$45$	&$54$	&$55$	\\
\midrule
-5 &$1$ &$0$ &$0$ &$0$ &$0$ &$0$ &$0$ &$0$\\
19 &$0$ &$0$ &$0$ &$-1$ &$-1$ &$1$ &$1$ &$-2$\\
43 &$-1$ &$1$ &$1$ &$2$ &$2$ &$0$ &$4$ &$4$\\
67 &$2$ &$0$ &$0$ &$7$ &$7$ &$14$ &$8$ &$8$\\
91 &$-4$ &$20$ &$20$ &$9$ &$9$ &$35$ &$62$ &$83$\\
115 &$5$ &$35$ &$35$ &$89$ &$89$ &$239$ &$251$ &$222$\\
139 &$20$ &$201$ &$201$ &$236$ &$236$ &$694$ &$881$ &$950$\\
163 &$36$ &$528$ &$528$ &$833$ &$833$ &$2323$ &$2702$ &$2665$\\
187 &$156$ &$1655$ &$1655$ &$2311$ &$2311$ &$6485$ &$7918$ &$8235$\\
211 &$413$ &$4263$ &$4263$ &$6423$ &$6423$ &$18051$ &$21397$ &$21498$\\
235 &$1020$ &$11382$ &$11382$ &$16087$ &$16087$ &$45358$ &$54939$ &$56379$\\
259 &$2455$ &$26994$ &$26994$ &$40021$ &$40021$ &$112341$ &$134076$ &$135889$\\
283 &$5879$ &$64617$ &$64617$ &$93003$ &$93003$ &$261869$ &$315210$ &$322155$\\
307 &$13169$ &$144938$ &$144938$ &$212050$ &$212050$ &$596095$ &$713920$ &$725468$\\
331 &$29087$ &$320286$ &$320286$ &$464053$ &$464053$ &$1305434$ &$1568719$ &$1600075$\\
355 &$61991$ &$680939$ &$680939$ &$993468$ &$993468$ &$2793599$ &$3348812$ &$3407543$\\
379 &$129169$ &$1422770$ &$1422770$ &$2064719$ &$2064719$ &$5808170$ &$6975047$ &$7109138$\\
403 &$262777$ &$2888505$ &$2888505$ &$4208325$ &$4208325$ &$11833971$ &$14193562$ &$14449220$\\
427 &$524286$ &$5767282$ &$5767282$ &$8379616$ &$8379616$ &$23570078$ &$28293712$ &$28827745$\\
451 &$1023996$ &$11264435$ &$11264435$ &$16396496$ &$16396496$ &$46112567$ &$55321988$ &$56332449$\\
475 &$1968752$ &$21659408$ &$21659408$ &$31488517$ &$31488517$ &$88564546$ &$106295853$ &$108283778$\\
499 &$3720441$ &$40917665$ &$40917665$ &$59539940$ &$59539940$ &$167451476$ &$200915361$ &$204609370$\\
523 &$6927917$ &$76215745$ &$76215745$ &$110825425$ &$110825425$ &$311704255$ &$374082188$ &$381045664$\\
547 &$12715863$ &$139866126$ &$139866126$ &$203487898$ &$203487898$ &$572297901$ &$686707842$ &$699375765$\\
571 &$23041098$ &$253459390$ &$253459390$ &$368608354$ &$368608354$ &$1036726053$ &$1244135671$ &$1267242033$\\
595 &$41225542$ &$453471967$ &$453471967$ &$659672220$ &$659672220$ &$1855310757$ &$2226288378$ &$2267426223$\\
619 &$72924892$ &$802191793$ &$802191793$ &$1166723003$ &$1166723003$ &$3281429787$ &$3937829833$ &$4010870330$\\
643 &$127575605$ &$1403299073$ &$1403299073$ &$2041300549$ &$2041300549$ &$5741128020$ &$6889198936$ &$7016622004$\\
667 &$220904792$ &$2429997813$ &$2429997813$ &$3534354858$ &$3534354858$ &$9940416134$ &$11928705034$ &$12149811849$\\
691 &$378758456$ &$4166294843$ &$4166294843$ &$6060311787$ &$6060311787$ &$17044567168$ &$20453213655$ &$20831707181$\\
715 &$643445607$ &$7077944811$ &$7077944811$ &$10294877321$ &$10294877321$ &$28954418447$ &$34745644272$ &$35389433549$\\
739 &$1083447396$ &$11917871022$ &$11917871022$ &$17335484390$ &$17335484390$ &$48755959253$ &$58506711305$ &$59589713385$\\
763 &$1809154794$ &$19900794975$ &$19900794975$ &$28946097935$ &$28946097935$ &$81411011737$ &$97693785348$ &$99503517664$\\
787 &$2996789379$ &$32964531410$ &$32964531410$ &$47949080384$ &$47949080384$ &$134856640526$ &$161827223016$ &$164823268101$\\
811 &$4926375742$ &$54190323007$ &$54190323007$ &$78821410398$ &$78821410398$ &$221685419135$ &$266023467395$ &$270950794599$\\
835 &$8039356619$ &$88432717409$ &$88432717409$ &$128630532359$ &$128630532359$ &$361773104706$ &$434126499529$ &$442164641760$\\
859 &$13028224735$ &$143310698119$ &$143310698119$ &$208450511320$ &$208450511320$ &$586267397566$ &$703522419579$ &$716552193823$\\
883 &$20971787728$ &$230689383931$ &$230689383931$ &$335549960886$ &$335549960886$ &$943733861399$ &$1132478688778$ &$1153448509388$\\
907 &$33542713964$ &$368970251977$ &$368970251977$ &$536681774077$ &$536681774077$ &$1509417984175$ &$1811304051250$ &$1844849248586$\\
931 &$53318392379$ &$586501744668$ &$586501744668$ &$853096286376$ &$853096286376$ &$2399332671290$ &$2879196063095$ &$2932511326216$\\
955 &$84251935429$ &$926772031772$ &$926772031772$ &$1348028436453$ &$1348028436453$ &$3791330807137$ &$4549600936442$ &$4633856811066$\\
979 &$132373064244$ &$1456102857556$ &$1456102857556$ &$2117972326676$ &$2117972326676$ &$5956796105864$ &$7148150369696$ &$7280518493936$\\
1003 &$206837818945$ &$2275216938197$ &$2275216938197$ &$3309400831213$ &$3309400831213$ &$9307691151506$ &$11169235537586$ &$11376079524539$\\
1027 &$321479780918$ &$3536276449763$ &$3536276449763$ &$5143681782673$ &$5143681782673$ &$14466603424870$ &$17359916463414$ &$17681388557332$\\
1051 &$497110820190$ &$5468220598525$ &$5468220598525$ &$7953766796046$ &$7953766796046$ &$22369971051653$ &$26843974790631$ &$27341095178372$\\
1075 &$764896637498$ &$8413860876702$ &$8413860876702$ &$12238353855121$ &$12238353855121$ &$34420367790300$ &$41304429463648$ &$42069314230101$\\
1099 &$1171318829840$ &$12884509771652$ &$12884509771652$ &$18741091700903$ &$18741091700903$ &$52709323491410$ &$63251202946890$ &$64422536456281$\\
1123 &$1785406214518$ &$19639465309775$ &$19639465309775$ &$28566511591466$ &$28566511591466$ &$80343309988087$ &$96411953800436$ &$98197341895753$\\
1147 &$2709277080802$ &$29802051397724$ &$29802051397724$ &$43348418062984$ &$43348418062984$ &$121917430520635$ &$146300938919330$ &$149010238333467$\\
\bottomrule
\end{tabular}
\end{center}
\end{scriptsize}
\end{sidewaystable}

\clearpage

\begin{sidewaystable}[ht]
\vspace{-8pt}
\begin{scriptsize}
\begin{center}
\caption{\small{Multiplicity generating functions $f^{(1)}_{\dpe,\chi}$, part 2}}\label{tab:mults:m12_2}\medskip
\begin{tabular}{c|@{ }r@{ }r@{ }r@{ }r@{ }r@{ }r@{ }r}\toprule
$24d$	&$55'$	&$55''$	&$66$	&$99$	&$120$	&$144$	&$176$	\\
\midrule
-5 &$0$ &$0$ &$0$ &$0$ &$0$ &$0$ &$0$\\
19 &$0$ &$0$ &$0$ &$0$ &$0$ &$2$ &$0$\\
43 &$3$ &$3$ &$1$ &$5$ &$2$ &$1$ &$9$\\
67 &$12$ &$12$ &$21$ &$21$ &$36$ &$35$ &$37$\\
91 &$65$ &$65$ &$72$ &$116$ &$130$ &$170$ &$207$\\
115 &$244$ &$244$ &$303$ &$450$ &$572$ &$668$ &$809$\\
139 &$910$ &$910$ &$1053$ &$1625$ &$1892$ &$2299$ &$2843$\\
163 &$2743$ &$2743$ &$3380$ &$4946$ &$6152$ &$7328$ &$8873$\\
187 &$8105$ &$8105$ &$9602$ &$14577$ &$17408$ &$21007$ &$25847$\\
211 &$21733$ &$21733$ &$26243$ &$39149$ &$47808$ &$57171$ &$69639$\\
235 &$56011$ &$56011$ &$66950$ &$100712$ &$121602$ &$146250$ &$178995$\\
259 &$136438$ &$136438$ &$164139$ &$245788$ &$298688$ &$357966$ &$437126$\\
283 &$321282$ &$321282$ &$384883$ &$578131$ &$699412$ &$839873$ &$1027282$\\
307 &$726800$ &$726800$ &$873253$ &$1308380$ &$1588170$ &$1904874$ &$2326881$\\
331 &$1598171$ &$1598171$ &$1916117$ &$2876429$ &$3483272$ &$4181523$ &$5112694$\\
355 &$3410318$ &$3410318$ &$4094710$ &$6139138$ &$7445828$ &$8932518$ &$10914961$\\
379 &$7105048$ &$7105048$ &$8522970$ &$12788159$ &$15494804$ &$18597231$ &$22733170$\\
403 &$14455051$ &$14455051$ &$17350401$ &$26020434$ &$31548542$ &$37853674$ &$46261017$\\
427 &$28819592$ &$28819592$ &$34577010$ &$51873536$ &$62864134$ &$75443098$ &$92215429$\\
451 &$56343865$ &$56343865$ &$67622177$ &$101421034$ &$122953390$ &$147535322$ &$180310064$\\
475 &$108267895$ &$108267895$ &$129908352$ &$194879528$ &$236191550$ &$283442660$ &$346445082$\\
499 &$204631232$ &$204631232$ &$245574876$ &$368340362$ &$446507564$ &$535790948$ &$654835854$\\
523 &$381015906$ &$381015906$ &$457195844$ &$685822008$ &$831254198$ &$997529608$ &$1219228018$\\
547 &$699415924$ &$699415924$ &$839330695$ &$1258957772$ &$1526071004$ &$1831253012$ &$2238164410$\\
571 &$1267187750$ &$1267187750$ &$1520582135$ &$2280927045$ &$2764674464$ &$3317651311$ &$4054954916$\\
595 &$2267499033$ &$2267499033$ &$2721057746$ &$4081511489$ &$4947403906$ &$5936828205$ &$7256056407$\\
619 &$4010773890$ &$4010773890$ &$4812849666$ &$7219375201$ &$8750602126$ &$10500800283$ &$12834400672$\\
643 &$7016749493$ &$7016749493$ &$8420202262$ &$12630174462$ &$15309503684$ &$18371299544$ &$22453696426$\\
667 &$12149643643$ &$12149643643$ &$14579440062$ &$21869323134$ &$26508010882$ &$31809749824$ &$38878728441$\\
691 &$20831927400$ &$20831927400$ &$24998485648$ &$37497515848$ &$45451874944$ &$54542074762$ &$66662346131$\\
715 &$35389146944$ &$35389146944$ &$42466747340$ &$63700406402$ &$77212161966$ &$92654819516$ &$113245032965$\\
739 &$59590085251$ &$59590085251$ &$71508403663$ &$107262225204$ &$130015413134$ &$156018202222$ &$190688575918$\\
763 &$99503036747$ &$99503036747$ &$119403255079$ &$179105374690$ &$217096655922$ &$260516372674$ &$318409337999$\\
787 &$164823887402$ &$164823887402$ &$197789160389$ &$296683119323$ &$359616877840$ &$431539751281$ &$527436921384$\\
811 &$270950001188$ &$270950001188$ &$325139371484$ &$487709837991$ &$591162203858$ &$709395286805$ &$867039376965$\\
835 &$442165654782$ &$442165654782$ &$530599589906$ &$795898389892$ &$964726901324$ &$1157671471836$ &$1414930915206$\\
859 &$716550903166$ &$716550903166$ &$859860053460$ &$1289791366536$ &$1563381441138$ &$1876058749736$ &$2292961837333$\\
883 &$1153450147408$ &$1153450147408$ &$1384141494283$ &$2076210582947$ &$2516621491560$ &$3019944492097$ &$3691041793857$\\
907 &$1844847178517$ &$1844847178517$ &$2213814945093$ &$3320724518828$ &$4025117339308$ &$4830142466496$ &$5903509327159$\\
931 &$2932513935701$ &$2932513935701$ &$3519018815116$ &$5278525605052$ &$6398216968080$ &$7677858253874$ &$9384046650468$\\
955 &$4633853529026$ &$4633853529026$ &$5560621623978$ &$8340935682923$ &$10110219936042$ &$12132266566371$ &$14828328689896$\\
979 &$7280522608177$ &$7280522608177$ &$8736630398824$ &$13104941535681$ &$15884784057350$ &$19061737583669$ &$23297675655094$\\
1003 &$11376074383581$ &$11376074383581$ &$13651285150689$ &$20476932857057$ &$24820516574124$ &$29784623972565$ &$36403433861254$\\
1027 &$17681394966175$ &$17681394966175$ &$21217679111305$ &$31826512201395$ &$38577600704766$ &$46293115745065$ &$56580469055649$\\
1051 &$27341087206947$ &$27341087206947$ &$32809298242028$ &$49213955409660$ &$59653266656898$ &$71583926371636$ &$87491472717522$\\
1075 &$42069324119087$ &$42069324119087$ &$50483196856446$ &$75724785382753$ &$91787634227288$ &$110145153118297$ &$134621845007103$\\
1099 &$64422524221421$ &$64422524221421$ &$77307019308615$ &$115960541117888$ &$140558212466154$ &$168669864792640$ &$206152067770248$\\
1123 &$98197356999292$ &$98197356999292$ &$117836840441407$ &$176755245669203$ &$214248806330278$ &$257098555519575$ &$314231554534832$\\
1147 &$149010219724356$ &$149010219724356$ &$178812248794508$ &$268218391773495$ &$325113172828678$ &$390135822209555$ &$476832688109889$\\
\bottomrule
\end{tabular}
\end{center}
\end{scriptsize}
\end{sidewaystable}

\clearpage

\begin{sidewaystable}[ht]
\vspace{-8pt}
\begin{scriptsize}
\begin{center}
\caption{\small{Multiplicity generating functions $f^{(1)}_{\dpe,\chi}$, part 3}}\label{tab:mults:m12_3}\medskip
\begin{tabular}{c|@{ }r@{ }r@{ }r@{ }r@{ }r}\toprule
$24d$	&$1$	&$11$	&$11'$	&$16$	&${16'}$		\\
\midrule
1171 &$4093401506990$ &$45027412291973$ &$45027412291973$ &$65494442717998$ &$65494442717998$\\
1195 &$6158685759395$ &$67745548890523$ &$67745548890523$ &$98538949769247$ &$98538949769247$\\
1219 &$9228239349225$ &$101510625677235$ &$101510625677235$ &$147651856584014$ &$147651856584014$\\
1243 &$13773021513812$ &$151503245506811$ &$151503245506811$ &$220368311165802$ &$220368311165802$\\
1267 &$20477101283532$ &$225248103749002$ &$225248103749002$ &$327633661514469$ &$327633661514469$\\
1291 &$30330817057006$ &$333638999586028$ &$333638999586028$ &$485293022305906$ &$485293022305906$\\
1315 &$44763186057643$ &$492395032232149$ &$492395032232149$ &$716211038247564$ &$716211038247564$\\
1339 &$65829766710070$ &$724127452040606$ &$724127452040606$ &$1053276194195726$ &$1053276194195726$\\
1363 &$96478169516795$ &$1061259841697878$ &$1061259841697878$ &$1543650799851638$ &$1543650799851638$\\
1387 &$140922592375703$ &$1550148543933051$ &$1550148543933051$ &$2254761371678042$ &$2254761371678042$\\
1411 &$205170323111458$ &$2256873521699226$ &$2256873521699226$ &$3282725299689321$ &$3282725299689321$\\
1435 &$297760377206316$ &$3275364187275260$ &$3275364187275260$ &$4764165877605326$ &$4764165877605326$\\
1459 &$430796502686742$ &$4738761484014951$ &$4738761484014951$ &$6892744232195987$ &$6892744232195987$\\
1483 &$621388898570251$ &$6835277940289662$ &$6835277940289662$ &$9942222151968754$ &$9942222151968754$\\
1507 &$893660568325681$ &$9830266182559709$ &$9830266182559709$ &$14298569361270602$ &$14298569361270602$\\
1531 &$1281533705334823$ &$14096870841750772$ &$14096870841750772$ &$20504538963932962$ &$20504538963932962$\\
1555 &$1832589066704092$ &$20158479636409529$ &$20158479636409529$ &$29321425454783979$ &$29321425454783979$\\
1579 &$2613400611330406$ &$28747406838047288$ &$28747406838047288$ &$41814409315423788$ &$41814409315423788$\\
1603 &$3716891022406435$ &$40885801111701610$ &$40885801111701610$ &$59470256913313545$ &$59470256913313545$\\
1627 &$5272453300096449$ &$57996986464667060$ &$57996986464667060$ &$84359252145641326$ &$84359252145641326$\\
1651 &$7459846110139069$ &$82058307013061490$ &$82058307013061490$ &$119357538538247375$ &$119357538538247375$\\
1675 &$10528232420636209$ &$115810556863115569$ &$115810556863115569$ &$168451717806641647$ &$168451717806641647$\\
1699 &$14822207710919004$ &$163044284544165030$ &$163044284544165030$ &$237155324477163395$ &$237155324477163395$\\
1723 &$20817315567358978$ &$228990471562943319$ &$228990471562943319$ &$333077047765997253$ &$333077047765997253$\\
1747 &$29168407625477909$ &$320852483499759068$ &$320852483499759068$ &$466694523558355316$ &$466694523558355316$\\
1771 &$40775370225169032$ &$448529072932153077$ &$448529072932153077$ &$652405921777788429$ &$652405921777788429$\\
1795 &$56872276713083413$ &$625595043298945690$ &$625595043298945690$ &$909956429556887554$ &$909956429556887554$\\
1819 &$79148095494043006$ &$870629051078933309$ &$870629051078933309$ &$1266369525369451301$ &$1266369525369451301$\\
1843 &$109909812683600080$ &$1209007938767976703$ &$1209007938767976703$ &$1758557005937509432$ &$1758557005937509432$\\
1867 &$152302485633902688$ &$1675327342846829765$ &$1675327342846829765$ &$2436839766599162279$ &$2436839766599162279$\\
1891 &$210605556865399279$ &$2316661124493547173$ &$2316661124493547173$ &$3369688914010001224$ &$3369688914010001224$\\
1915 &$290631176879715729$ &$3196942946893625371$ &$3196942946893625371$ &$4650098825202784367$ &$4650098825202784367$\\
1939 &$400258719166874651$ &$4402845909393277271$ &$4402845909393277271$ &$6404139512372754278$ &$6404139512372754278$\\
1963 &$550150869077258932$ &$6051659561542349276$ &$6051659561542349276$ &$8802413898544613901$ &$8802413898544613901$\\
1987 &$754711357815895642$ &$8301824934007779454$ &$8301824934007779454$ &$12075381732915932852$ &$12075381732915932852$\\
2011 &$1033363835278555220$ &$11367002190347074914$ &$11367002190347074914$ &$16533821355238048767$ &$16533821355238048767$\\
2035 &$1412256805146747435$ &$15534824853948167213$ &$15534824853948167213$ &$22596108893118543439$ &$22596108893118543439$\\
2059 &$1926533042028023191$ &$21191863465442002954$ &$21191863465442002954$ &$30824528659900088909$ &$30824528659900088909$\\
2083 &$2623345678439652253$ &$28856802459153067204$ &$28856802459153067204$ &$41973530869649280580$ &$41973530869649280580$\\
2107 &$3565860601400413225$ &$39224466619703622922$ &$39224466619703622922$ &$57053769605360276336$ &$57053769605360276336$\\
2131 &$4838559710474916833$ &$53224156810243842644$ &$53224156810243842644$ &$77416955387499037845$ &$77416955387499037845$\\
2155 &$6554257603147504112$ &$72096833640381599416$ &$72096833640381599416$ &$104868121627158155768$ &$104868121627158155768$\\
2179 &$8863371782121759020$ &$97497089596648487456$ &$97497089596648487456$ &$141813948540918808370$ &$141813948540918808370$\\
2203 &$11966152827447202656$ &$131627681109732904436$ &$131627681109732904436$ &$191458445207879722486$ &$191458445207879722486$\\
2227 &$16128796948623520740$ &$177416766425737654861$ &$177416766425737654861$ &$258060751214233038172$ &$258060751214233038172$\\
2251 &$21704644236610653731$ &$238751086613304015784$ &$238751086613304015784$ &$347274307743692860669$ &$347274307743692860669$\\
2275 &$29162029847290462732$ &$320782328307975295104$ &$320782328307975295104$ &$466592477605502723278$ &$466592477605502723278$\\
2299 &$39120827378017906356$ &$430329101172286761258$ &$430329101172286761258$ &$625933237991633696014$ &$625933237991633696014$\\
\bottomrule
\end{tabular}
\end{center}
\end{scriptsize}
\end{sidewaystable}

\clearpage

\begin{sidewaystable}[ht]
\vspace{-8pt}
\begin{scriptsize}
\begin{center}
\caption{\small{Multiplicity generating functions $f^{(1)}_{\dpe,\chi}$, part 4}}\label{tab:mults:m12_4}\medskip
\begin{tabular}{c|@{ }r@{ }r@{ }r@{ }r@{ }r}\toprule
$24d$	&$45$	&$54$	&$55$	&$55'$	&$55''$		\\
\midrule
1171 &$184203114462622$ &$221043710022343$ &$225137084071984$ &$225137106948035$ &$225137106948035$\\
1195 &$277140803111755$ &$332568997319530$ &$338727716741557$ &$338727688687159$ &$338727688687159$\\
1219 &$415270838158915$ &$498324964525296$ &$507553162674521$ &$507553197014225$ &$507553197014225$\\
1243 &$619785885692343$ &$743743113342304$ &$757516185209007$ &$757516143250948$ &$757516143250948$\\
1267 &$921469660027838$ &$1105763530523634$ &$1126240570403383$ &$1126240621565028$ &$1126240621565028$\\
1291 &$1364886640954779$ &$1637864043787277$ &$1668194935551870$ &$1668194873292078$ &$1668194873292078$\\
1315 &$2014343526237095$ &$2417212140959443$ &$2461975236268978$ &$2461975311902653$ &$2461975311902653$\\
1339 &$2962339318825550$ &$3554807292476658$ &$3620637169265217$ &$3620637077537819$ &$3620637077537819$\\
1363 &$4341517847042590$ &$5209821283094568$ &$5306299319337420$ &$5306299430386979$ &$5306299430386979$\\
1387 &$6341516391505723$ &$7609819831217510$ &$7750742584630229$ &$7750742450426761$ &$7750742450426761$\\
1411 &$9232664864481335$ &$11079197642782316$ &$11284367771598830$ &$11284367933524970$ &$11284367933524970$\\
1435 &$13399216579858759$ &$16079060129761174$ &$16376820741075612$ &$16376820545992293$ &$16376820545992293$\\
1459 &$19385843094557620$ &$23263011432427915$ &$23693807653507039$ &$23693807888163762$ &$23693807888163762$\\
1483 &$27962499872265004$ &$33555000184510544$ &$34176389421242631$ &$34176389139433773$ &$34176389139433773$\\
1507 &$40214726244543434$ &$48257671087653672$ &$49151331250463847$ &$49151331588413804$ &$49151331588413804$\\
1531 &$57669015937475566$ &$69202819611374765$ &$70484353802379119$ &$70484353397662854$ &$70484353397662854$\\
1555 &$82466508969735861$ &$98959810182387002$ &$100792398668290829$ &$100792399152265772$ &$100792399152265772$\\
1579 &$117603026344789245$ &$141123632306950527$ &$143737033611788997$ &$143737033033854903$ &$143737033033854903$\\
1603 &$167260097396709360$ &$200712116050021046$ &$204429006245393365$ &$204429006934618721$ &$204429006934618721$\\
1627 &$237260396863613020$ &$284712477220607098$ &$289984931505798416$ &$289984930684900175$ &$289984930684900175$\\
1651 &$335693076895668315$ &$402831691102628332$ &$410291536040979311$ &$410291537017434228$ &$410291537017434228$\\
1675 &$473770456621652456$ &$568524549339514438$ &$579052783152120837$ &$579052781992128638$ &$579052781992128638$\\
1699 &$666999349746211386$ &$800399218042669045$ &$815221424102015049$ &$815221425478366554$ &$815221425478366554$\\
1723 &$936779197250978703$ &$1124135038657868408$ &$1144952356182632432$ &$1144952354551480412$ &$1144952354551480412$\\
1747 &$1312578347025836554$ &$1575094014116214816$ &$1604262419424649864$ &$1604262421355471132$ &$1604262421355471132$\\
1771 &$1834891655568344765$ &$2201869989419897616$ &$2242645362384455501$ &$2242645360101610419$ &$2242645360101610419$\\
1795 &$2559252457456586760$ &$3071102945711666490$ &$3127975219189577966$ &$3127975221885611857$ &$3127975221885611857$\\
1819 &$3561664290897644899$ &$4273997152896785912$ &$4353145252207517405$ &$4353145249026968359$ &$4353145249026968359$\\
1843 &$4945941578258894303$ &$5935129889410954110$ &$6045039697596848087$ &$6045039701344867210$ &$6045039701344867210$\\
1867 &$6853611844665722902$ &$8224334218891965090$ &$8376636709820185138$ &$8376636705408223928$ &$8376636705408223928$\\
1891 &$9477250069357366762$ &$11372700077007117472$ &$11583305627646850394$ &$11583305632834988246$ &$11583305632834988246$\\
1915 &$13078402947401857383$ &$15694083544192819487$ &$15984714728386214908$ &$15984714722291519041$ &$15984714722291519041$\\
1939 &$18011642376763874898$ &$21613970843532086552$ &$22014229554115942440$ &$22014229561268362549$ &$22014229561268362549$\\
1963 &$24756789091754815561$ &$29708146920173901712$ &$30258297799313493059$ &$30258297790928171326$ &$30258297790928171326$\\
1987 &$33962011121364327166$ &$40754413333847397958$ &$41509124679877898687$ &$41509124689699169994$ &$41509124689699169994$\\
2011 &$46501372564484998594$ &$55801647091170275277$ &$56835010940239663338$ &$56835010928747353467$ &$56835010928747353467$\\
2035 &$63551556258539592867$ &$76261867494133421282$ &$77674124283157975065$ &$77674124296593010152$ &$77674124296593010152$\\
2059 &$86693986859883218954$ &$104032784250684215406$ &$105959317311542087994$ &$105959317295850490724$ &$105959317295850490724$\\
2083 &$118050555566317547234$ &$141660666657604661458$ &$144284012314071615021$ &$144284012332382347583$ &$144284012332382347583$\\
2107 &$160463727020416092994$ &$192556472450127762662$ &$196122333077146178381$ &$196122333055797906898$ &$196122333055797906898$\\
2131 &$217735187021112290780$ &$261282224395485799571$ &$266120784076119006487$ &$266120784100986919412$ &$266120784100986919412$\\
2155 &$294941592083627118072$ &$353929910535079510346$ &$360484168172958312466$ &$360484168144015486852$ &$360484168144015486852$\\
2179 &$398851730262924276402$ &$478622076275134635183$ &$487485448016868002166$ &$487485448050525203783$ &$487485448050525203783$\\
2203 &$538476877156922474149$ &$646172252635225208162$ &$658138405509601252323$ &$658138405470493978887$ &$658138405470493978887$\\
2227 &$725795862778692020557$ &$870955035279941337060$ &$887083832174081304741$ &$887083832219484026698$ &$887083832219484026698$\\
2251 &$976708990542309495285$ &$1172050788713993804393$ &$1193755433013807220005$ &$1193755432961138141154$ &$1193755432961138141154$\\
2275 &$1312291343250192222140$ &$1574749611826956036768$ &$1603911641600986492613$ &$1603911641662036771525$ &$1603911641662036771525$\\
2299 &$1760437231869163874840$ &$2112524678327840972437$ &$2151645505790711677538$ &$2151645505720001135999$ &$2151645505720001135999$\\
\bottomrule
\end{tabular}
\end{center}
\end{scriptsize}
\end{sidewaystable}

\clearpage

\begin{sidewaystable}[ht]
\vspace{-8pt}
\begin{scriptsize}
\begin{center}
\caption{\small{Multiplicity generating functions $f^{(1)}_{\dpe,\chi}$, part 5}}\label{tab:mults:m12_5}\medskip
\begin{tabular}{c|@{ }r@{ }r@{ }r@{ }r@{ }r}\toprule
$24d$	&$66$	&$99$	&$120$	&$144$	&$176$		\\
\midrule
1171 &$270164546681156$ &$405246797027805$ &$491208274997242$ &$589449911767971$ &$720438760605286$\\
1195 &$406473203907733$ &$609709834093878$ &$739042178780400$ &$886850636998028$ &$1083928581424780$\\
1219 &$609063863910983$ &$913595761484332$ &$1107388855916070$ &$1328866599517240$ &$1624170257738622$\\
1243 &$909019338414137$ &$1363529049387265$ &$1652762418194392$ &$1983314935496409$ &$2424051624958728$\\
1267 &$1351488786694107$ &$2027233129147441$ &$2457252358138100$ &$2948702788864913$ &$3603970030037952$\\
1291 &$2001833798159548$ &$3002750759449688$ &$3639697792134294$ &$4367637400215692$ &$5338223544445360$\\
1315 &$2954370434912214$ &$4431555576448250$ &$5371582636389448$ &$6445899103301566$ &$7878321058779942$\\
1339 &$4344764419512895$ &$6517146721381044$ &$7899571638612220$ &$9479486039749018$ &$11586038574903016$\\
1363 &$6367559405323185$ &$9551338996863518$ &$11577380777243530$ &$13892856843630246$ &$16980158265658175$\\
1387 &$9300890833306805$ &$13951336383760152$ &$16910710557156234$ &$20292852776170360$ &$24802375734252748$\\
1411 &$13541241649561517$ &$20311862312968942$ &$24620439421902434$ &$29544527176781278$ &$36109977517072127$\\
1435 &$19652184499179137$ &$29478276943727693$ &$35731244472881150$ &$42877493523182575$ &$52405825590524819$\\
1459 &$28432569653735348$ &$42648854245377598$ &$51695581273866940$ &$62034697341246448$ &$75820185430196005$\\
1483 &$41011666741538202$ &$61517500394937483$ &$74566666700591406$ &$89480000266235449$ &$109364445021070861$\\
1507 &$58981598176550861$ &$88472396926681358$ &$107239269534589732$ &$128687123170645012$ &$157284261352456668$\\
1531 &$84581223753771591$ &$126871836034520310$ &$153784043041385876$ &$184540851973883398$ &$225549930549289886$\\
1555 &$120950879369401488$ &$181426318571322984$ &$219910689938939870$ &$263892827539670898$ &$322535677674888309$\\
1579 &$172484439178360861$ &$258726659345247778$ &$313608071022950170$ &$376329685689204628$ &$459958505244752664$\\
1603 &$245314808873442099$ &$367972212619696564$ &$446026925474876674$ &$535232310019095576$ &$654172822742952752$\\
1627 &$347981916164512698$ &$521972875069314411$ &$632694392728415992$ &$759233271930954787$ &$927951777535669872$\\
1651 &$492349845202179646$ &$738524766826508033$ &$895181537085930972$ &$1074217843720999295$ &$1312932919235178135$\\
1675 &$694863337463226605$ &$1042295007353137208$ &$1263387885874730544$ &$1516065463978597964$ &$1852968901447872354$\\
1699 &$978265711674229209$ &$1467398566137307066$ &$1778664930818030794$ &$2134397915880732844$ &$2608708562632879662$\\
1723 &$1373942824157073997$ &$2060914237866252831$ &$2498077861510042094$ &$2997693435115570067$ &$3663847533257402807$\\
1747 &$1925114907172125283$ &$2887672358825009115$ &$3500208922832935414$ &$4200250705856191083$ &$5133639749883610819$\\
1771 &$2691174430294308510$ &$4036761647727631051$ &$4893044417887901864$ &$5871653303292074233$ &$7176465150500213001$\\
1795 &$3753570268419902966$ &$5630355399933081775$ &$6824673216288466950$ &$8189607857387351047$ &$10009520712187446446$\\
1819 &$5223774296289292241$ &$7835661447611118530$ &$9497771446641673624$ &$11397325738516222932$ &$13930064794343989332$\\
1843 &$7254047644610345604$ &$10881071463172141658$ &$13189177537019635416$ &$15827013041425604314$ &$19344127047303818283$\\
1867 &$10051964042960644068$ &$15077946068852005519$ &$18276298258323001986$ &$21931557913514503889$ &$26805237453771981804$\\
1891 &$13899966763554443537$ &$20849950140138817382$ &$25272666844712164358$ &$30327200209506516810$ &$37066578029225470889$\\
1915 &$19181657661871572164$ &$28772486498908357454$ &$34875741201188675430$ &$41850889446302706172$ &$51151087106459775710$\\
1939 &$26417075479244374467$ &$39625613211712943497$ &$48031046328498284852$ &$57637255588472376133$ &$70445534601774047231$\\
1963 &$36309957342408061267$ &$54464936021991049526$ &$66018104255873659084$ &$79221725113760100204$ &$96826552924265854344$\\
1987 &$49810949635492675761$ &$74716424443426330769$ &$90565362977197226288$ &$108678435564780334215$ &$132829199014898120022$\\
2011 &$68202013105303745330$ &$102303019669446105608$ &$124003660187280231114$ &$148804392233925001344$ &$181872034962788778272$\\
2035 &$93208949166662983780$ &$139813423736550819498$ &$169470816671544273222$ &$203364979995109897828$ &$248557197759850986121$\\
2059 &$127151180742462470212$ &$190726771129397283986$ &$231183964980593334826$ &$277420757989266397324$ &$339069815334173156022$\\
2083 &$173140814813508567478$ &$259711222201949509299$ &$314801481485761998530$ &$377761777768258741081$ &$461708839478260064519$\\
2107 &$235346799649883728800$ &$353020199496161859438$ &$427903272082932960552$ &$513483926516604492932$ &$627591465761482246451$\\
2131 &$319344940941071942661$ &$479017411386756381538$ &$580627165356451864750$ &$696752598407849494756$ &$851586509143058978448$\\
2155 &$432581001749665552605$ &$648871502653437926113$ &$786510912261590705900$ &$943813094737053749501$ &$1153549338037678618915$\\
2179 &$584982537687562696957$ &$877473806497670520373$ &$1063604613989623332136$ &$1276325536760630740267$ &$1559953433788616761379$\\
2203 &$789766086533298317643$ &$1184649129839076522389$ &$1435938339137237812658$ &$1723126006995973066501$ &$2106042897474304114574$\\
2227 &$1064500598699704522077$ &$1596750898004149803311$ &$1935455634015967754356$ &$2322546760782826588291$ &$2838668263138648397534$\\
2251 &$1432506519511239076275$ &$2148759779319506061861$ &$2604557308183097498046$ &$3125468769861864014025$ &$3820017385433520294132$\\
2275 &$1924693970043272902043$ &$2887040955003888251120$ &$3499443581919069498760$ &$4199332298254045573934$ &$5132517253367370460268$\\
2299 &$2581974606807435465725$ &$3872961910281857647683$ &$4694499285078708148956$ &$5633399142151001036453$ &$6885265618247405253843$\\
\bottomrule
\end{tabular}
\end{center}
\end{scriptsize}
\end{sidewaystable}

\end{document}